\RequirePackage{ifpdf,xcolor,tikz}
\documentclass[hyper,letterpaper]{JHEP3}
\pdfoutput=1
\usepackage{amsmath,amssymb,amsfonts,bm,empheq}
\usepackage{cite}
\usepackage{subcaption}
\usepackage{nicefrac}
\usepackage[enableskew,vcentermath]{youngtab}
\usepackage{epstopdf}
\usepackage{url}
\usepackage{float}
\usepackage{slashed,framed,xcolor,tikz}
\def\printname#1{
        \if\draft y
                \smash{\makebox[0pt]{\hspace{-0.5in}
                        \raisebox{8pt}{\tt\tiny #1}}}
        \fi
}

\newlength{\standardunitlength}
\catcode`\@=11
\long\def\@makecaption#1#2{%
     \vskip 10pt

\setbox\@tempboxa\hbox{%\ifvoid\tinybox\else\box\tinybox\fi
       \small\sf{\bfcaptionfont #1. }\ignorespaces #2}%
     \ifdim \wd\@tempboxa >\captionwidth {%
         \rightskip=\@captionmargin\leftskip=\@captionmargin
         \unhbox\@tempboxa\par}%
       \else
         \hbox to\hsize{\hfil\box\@tempboxa\hfil}%
     \fi}
\font\bfcaptionfont=cmssbx10 scaled \magstephalf
\newdimen\@captionmargin\@captionmargin=2\parindent
\newdimen\captionwidth\captionwidth=\hsize
\catcode`\@=12

\newdimen\tableauside\tableauside=1.0ex
\newdimen\tableaurule\tableaurule=0.4pt
\newdimen\tableaustep
\def\phantomhrule#1{\hbox{\vbox to0pt{\hrule height\tableaurule width#1\vss}}}
\def\phantomvrule#1{\vbox{\hbox to0pt{\vrule width\tableaurule height#1\hss}}}
\def\sqr{\vbox{%
\phantomhrule\tableaustep
\hbox{\phantomvrule\tableaustep\kern\tableaustep\phantomvrule\tableaustep}%
\hbox{\vbox{\phantomhrule\tableauside}\kern-\tableaurule}}}
\def\squares#1{\hbox{\count0=#1\noindent\loop\sqr
\advance\count0 by-1 \ifnum\count0>0\repeat}}
\def\tableau#1{\vcenter{\offinterlineskip
\tableaustep=\tableauside\advance\tableaustep by-\tableaurule
\kern\normallineskip\hbox
    {\kern\normallineskip\vbox
      {\gettableau#1 0 }%
     \kern\normallineskip\kern\tableaurule}%
  \kern\normallineskip\kern\tableaurule}}
\def\gettableau#1 {\ifnum#1=0\let\next=\null\else
  \squares{#1}\let\next=\gettableau\fi\next}

\newcommand{\bea}{\begin{eqnarray}}
\newcommand{\eea}{\end{eqnarray}}
\newcommand{\be}{\begin{equation}}
\newcommand{\ee}{\end{equation}}

\renewcommand{\bar}{\overline}
\renewcommand{\hat}{\widehat}

\newcommand\catalannumber[3]{
  \fill[white]  (#1) rectangle +(#2,#2);
  \fill[fill=gray!25]
  (#1)
  \foreach \dir in {#3}{
    \ifnum\dir=0
    -- ++(0,1)
    \else
    -- ++(1,0)
    \fi
  } |- (#1);
  \draw[help lines] (#1) grid +(#2,#2);
  \draw[dashed] (#1) -- +(#2,#2);
  \coordinate (prev) at (#1);
  \foreach \dir in {#3}{
    \ifnum\dir=0
    \coordinate (dep) at (0,1);
    \else
    \coordinate (dep) at (1,0);
    \fi
    \draw[line width=2pt,-stealth] (prev) -- ++(dep) coordinate (prev);
  };
}

\usetikzlibrary{calc}

\usepackage{mathtools}
\def\multiset#1#2{\ensuremath{\left(\kern-.1em\left(\genfrac{}{}{0pt}{}{#1}{#2}\right)\kern-.1em\right)}}

\newcommand*\pFqskip{8mu}

\catcode`,\active
\newcommand*\pFq{\begingroup
  \catcode`\,\active
  \def ,{\mskip\pFqskip\relax}%
  \dopFq
}
\catcode`\,12
\def\dopFq#1#2#3#4#5{%
  {}_{#1}F_{#2}\biggl[\genfrac..{0pt}{}{#3}{#4};#5\biggr]%
  \endgroup
}

\catcode`,\active
\newcommand*\pphiq{\begingroup
  \catcode`\,\active
  \def ,{\mskip\pFqskip\relax}%
  \dopphiq
}
\catcode`\,12
\def\dopphiq#1#2#3#4#5#6{%
  {}_{#1}\phi_{#2}\biggl[\genfrac..{0pt}{}{#3}{#4};#5,#6\biggr]%
  \endgroup
}

\title{Topological strings, strips and quivers}

\author{Mi{\l}osz Panfil$^{1}$ and Piotr Su{\l}kowski$^{1,2}$
\\
$^1$ Faculty of Physics, University of Warsaw, ul. Pasteura 5, 02-093 Warsaw, Poland \\
$^2$ Walter Burke Institute for Theoretical Physics, California Institute of Technology, Pasadena, CA 91125, USA 
}

\abstract{We find a direct relation between quiver representation theory and open topological string theory on a class of toric Calabi-Yau manifolds without compact four-cycles, also referred to as strip geometries. We show that various quantities that characterize open topological string theory on these manifolds, such as partition functions, Gromov-Witten invariants, or open BPS invariants, can be expressed in terms of characteristics of the moduli space of representations of the corresponding quiver. This has various deep consequences; in particular, expressing open BPS invariants in terms of motivic Donaldson-Thomas invariants, immediately proves integrality of the former ones.  Taking advantage of the relation to quivers we also derive explicit expressions for classical open BPS invariants for an arbitrary strip geometry, which lead to a large set of number theoretic integrality statements. Furthermore, for a specific framing, open topological string partition functions for strip geometries take form of generalized $q$-hypergeometric functions, which leads to a novel representation of these functions in terms of quantum dilogarithms and integral invariants. We also study quantum curves and A-polynomials associated to quivers, various limits thereof, and their specializations relevant for strip geometries. The relation between toric manifolds and quivers can be regarded as a generalization of the knots-quivers correspondence to more general Calabi-Yau geometries. 
\\
\\
\\
\\
\\
\\
\\ 
\\
\\
\\
\\
{\tt CALT-2018-047}}

\begin{document}

\tableofcontents

%%%%%%%%%%%%%%%%%%%%%%%%%%%%%%%%%%%%%%%%%%%%%%%%%%%%%%%%%%%%%%%%%%%%%

\newpage

\section{Introduction}    \label{sec-intro}

Topological string theory provides an interesting playground that enables exact computations of quantum amplitudes and analysis of various phenomena in a simplified setting. It is related to various other physical systems, such as supersymmetric gauge theories, surface operators, vortex counting, two-dimensional conformal field theory, BPS states, etc. Among various techniques to compute topological string amplitudes, a very powerful one relies on links with Chern-Simons theory. In particular, such links give rise to the topological vertex formalism \cite{AKMV,ADKMV}, which enables computation of closed and open topological string amplitudes for a large class of toric Calabi-Yau threefolds.

The relation between topological string theory and Chern-Simons theory also results in the connection with knot theory. On one hand, it is known that knot invariants can be computed as expectation values of Wilson loops in Chern-Simons theory \cite{Witten_Jones}. On the other hand, such Wilson loop configurations can be realized in string theory by choosing as a Calabi-Yau space the deformed conifold $T^*S^3$, and engineering a knot as an intersection of the base $S^3$ with an additional lagrangian brane \cite{Witten:1992fb,OoguriV}. In this case brane amplitudes turn out to reproduce Chern-Simons amplitudes associated to the engineered knot. Furthermore, upon the conifold transition this system is related to a lagrangian brane in the resolved conifold geometry, and in consequence various knot invariants can be expressed in terms of topological string amplitudes in the resolved conifold. Moreover, embedding this system in M-theory gives rise to new knot invariants, referred to as Labastida-Mari{\~n}o-Ooguri-Vafa (LMOV) invariants or simply open BPS invariants, which count BPS states of M5 and M2-branes, and thus are conjecturally integer \cite{OoguriV,Labastida:2000zp,Labastida:2000yw,Labastida:2001ts,Ramadevi_Sarkar,Mironov:2017hde}. 

Recently, motivated by such string theory considerations, knot invariants were related to yet another branch of mathematics, namely to quiver representation theory. This relation is referred to as the knots-quivers correspondence \cite{Kucharski:2017poe,Kucharski:2017ogk}; it states that to a given knot one can associate a quiver, so that various knot invariants are expressed in terms of quantities that characterize the moduli space of representations of this corresponding quiver. In particular, LMOV invariants for symmetric representations are expressed as integral linear combinations of motivic Donaldson-Thomas invariants associated to the quiver. The fact that the latter invariants are proven to be integer, proves the long sought after integrality of LMOV invariants, at least for symmetric representations. For related work and other aspects of knots-quiver correspondence see \cite{Kucharski:2016rlb,Luo:2016oza,Stosic:2017wno,Panfil:2018sis,EKL}.

In order to engineer more complicated knots in the above string theory setup, one needs to consider more complicated lagrangian branes in the the resolved conifold, which is one of the simplest Calabi-Yau manifold. The main idea in this paper is to consider the opposite situation -- we focus on simple examples of Aganagic-Vafa branes \cite{AV-discs,AKV-framing}, however embedded in more complicated toric Calabi-Yau manifolds. The manifolds that we consider do not have four-cycles and are referred to as strip geometries or generalized conifolds. We show that partition functions for branes in such manifolds can be also expressed as motivic generating functions of corresponding quivers, which we explicitly identify. This has various interesting consequences. Among others, it immediately leads to the proof of integrality of open BPS invariants associated to such brane systems, which are also referred to as Ooguri-Vafa invariants. Taking advantage of the relation to quivers we also derive explicit expressions for classical open BPS invariants for manifolds under consideration. More generally, it follows that various quantities that characterize topological strings can be reformulated as invariants of moduli spaces of quiver representations. One important consequence of this relation is the identification of the algebra of BPS states \cite{Harvey:1996gc} on the topological string side with the cohomological Hall algebra introduced in \cite{Kontsevich:2010px}. Moreover, the moduli space of representations of the corresponding quiver itself can be regarded as a new topological string invariant, thereby providing a novel categorification of topological string theory. Furthermore, various operations on both sides of the correspondence are matched, for example a change of framing of a brane by some number corresponds to adding the same number of loops at one particular vertex of the quiver. The identification of quivers corresponding to toric manifolds can be regarded as the generalization of the knots-quivers correspondence to more general toric Calabi-Yau manifolds.

It is also important to understand the meaning of quivers and the reason why they appear. Our results imply that vertices in these quivers should have a natural interpretation as corresponding to discs that represent open BPS states associated to a strip geometry, one of which is attached to the brane and other ones wrap hemispheres of all local $\mathbb{P}^1$'s (for resolved conifold such discs correspond to its two non-zero BPS invariants). A similar interpretation of quivers' vertices in the context of knots-quivers correspondence is presented in \cite{EKL}. On the other hand, from the physics perspective, analogously as in \cite{Kucharski:2017poe,Kucharski:2017ogk}, we postulate that the resulting quivers are associated to the effective supersymmetric quantum mechanics describing BPS states in the engineered brane systems; it would be nice to derive such a description more directly. 

It is also worth recalling that strip geometries that we consider in this paper have important properties and various applications. Their toric diagrams can be constructed as dual diagrams to a triangulation of a rectangular strip. Topological string partition functions for this class of geometries can be computed using the rules of the ``vertex on a strip'' \cite{Iqbal:2004ne}, which follow from the topological vertex formalism \cite{AKMV}. Strip geometries are a large class of manifolds, the simplest examples being $\mathbb{C}^3$, the resolved conifold, and resolutions of $\mathbb{C}^3/\mathbb{Z}_N$ orbifolds. In particular, the basic Aganagic-Vafa lagrangian brane in the resolved conifold engineers the unknot, and open topological string amplitudes in this case reproduce its colored HOMFLY-PT polynomials, so that the corresponding quiver provides a simple example of the knots-quivers correspondence. On the other hand, resolutions of $\mathbb{C}^3/\mathbb{Z}_N$ orbifolds and other examplse of strip geometries provide building blocks crucial for engineering of four-dimensional supersymmetric gauge theories, and lagrangian branes in such geometries engineer surface operators, as well as vortex counting in two normal spacetime dimensions. All these relations to other systems provide additional important motivations to study topological strings on strip geometries, and thus the corresponding quivers that we identify in this work.

Apart from revealing the correspondence to quivers, in this paper we present several other related, albeit at the same time independent results. First, we show that partition functions for branes in strip geometries take form of generalized $q$-hypergeometric functions $_r\phi_s$. General properties of these functions are studied e.g. in \cite{gasper2004basic}. This immediately leads to a non-trivial statement, that each generalized $q$-hypergeometric function is encoded in a series of integral BPS invariants, or motivic Donaldson-Thomas invariants of the corresponding quiver, and each such function can be written as the product of quantum dilogarithms. Furthermore these functions, in appropriate limit, reduce to (ordinary) generalized hypergeometric functions $_r F_s$. Therefore the information about each generalized hypergeometric function $_r F_s$ is also encoded in a set of integral BPS invariants, or motivic Donaldson-Thomas invariants for the corresponding quiver. Note that brane partition functions in the form of $q$-hypergeometric functions $_{r+1}\phi_r$, for a special class of strip geometries with all $\mathbb{P}^1$'s of $(-1,-1)$ type, were derived in \cite{Bonelli:2011fq,Kimura:2018kaf}, however it seems that the relation between arbitrary strip geometries and all $q$-hypergeometric functions $_r\phi_s$ has not been discussed before. 

Second, the form of brane partition functions motivates us to introduce a novel classical limit of quiver generating functions, that we refer to as the partial limit. We derive explicit formulas for coefficients of generating functions in this partial limit. These results generalize the explicit expressions for the ordinary classical generating functions derived in \cite{Panfil:2018sis}. Specializing these results to quivers associated to strip geometries we find explicit formulas for functions that satisfy mirror curve equations for an arbitrary strip geometry, and we also derive explicit expressions for classical open Ooguri-Vafa BPS invariants for an arbitrary strip geometry. This also means that mirror curves for strip geometries provide a large class of examples of algebraic equations satisfied by generating functions of Donaldson-Thomas invariants, illustrating the ideas in \cite{Mainiero:2016xaj}.

Third, we associate to quivers quantum curves, or A-polynomials, and analyze their properties and various limits. In particular we show that such A-polynomials, for quivers associated to strip geometries, are identified with quantum and classical mirror curves for such geometries. This enables us to study properties of mirror curves by taking advantage of tools of quiver representation theory. Note that various classes of curves associated to quivers, analogous to A-polynomials, are also studied in \cite{Panfil:2018sis,Smolinski-mgr,EKL}.

\subsection{A brief quantitative summary...}

Before starting detailed analysis, it may be of advantage to summarize main quantitative results of this work. Consider an arbitrary strip geometry, as shown in fig. \ref{fig-strip}, and the Aganagic-Vafa brane with a modulus $x$ in such geometry, in framing $f$, as shown in fig. \ref{fig-strip-2}. We first show that the partition function for such a brane takes form (\ref{psi-hyper})
\be
\psi_f(x) = \sum_{n=0}^{\infty} \big((-1)^{n} q^{n(n-1)/2}  \big)^{f+1}  \frac{ x ^n}{(q;q)_n} \frac{(\alpha_1;q)_n (\alpha_2;q)_n\cdots (\alpha_r;q)_n}{(\beta_1;q)_n (\beta_2;q)_n\cdots (\beta_s;q)_n},   \label{psi-f-intro}
\ee
where closed K{\"a}hler parameters $Q_k$ are encoded in variables $\alpha_i$ and $\beta_j$, and $(\alpha;q)_n$ is the $q$-Pochhammer symbol. Second, we show that the quantum mirror curve that annihilates this partition function, $\widehat{A}(\hat x,\hat y) \psi_f(x) = 0$, takes form (\ref{A-hat-psi-f})
\be
\widehat{A}(\hat x,\hat y) = (1 - \hat y)\prod_{j=1}^s (1 - q^{-1} \beta_j \hat y) + (-1)^{f} \hat{x}\,  \Big(\prod_{j=1}^r (1-\alpha_j\hat y) \Big) \hat{y}^{f+1},    \label{A-hat-intro}
\ee
with $\hat y$ defined such that $\hat{y}\psi(x)=\psi(qx)$, so that $\hat{y}\hat{x}=q\hat{x}\hat{y}$. Interestingly, for $f=s-r$ the brane partition function $\psi_f(x)$ reduces to the generalized $q$-hypergeometric function
\be
\psi_{s-r}(x) = \pphiq{r}{s}{\alpha_1,\alpha_2, \dots, \alpha_r}{\beta_1, \beta_2, \dots, \beta_r}{q}{x},
\ee
and the above quantum mirror curve equation takes form of the generalized $q$-hypergeometric equation. These results have also two interesting limits that we discuss. First, for $q\to 1$ the operator $\widehat{A}(\hat x,\hat y)$ reduces to the classical mirror curve equation $A(x,y)=0$, whose solution for $y=\sum_i c_i x^i$ we determine explicitly for an arbitrary strip geometry in (\ref{y-c_i}), by taking advantage of the relation to quivers. Second, setting $x \to (q-1)^{1+s-r} x, \alpha_i= q^{a_i}, \beta_j = q^{b_j}$, and then taking $q\to 1$ limit, the partition function $\psi_{s-r}(x)$ reduces to the ordinary generalized hypergeometric function (\ref{tilde-psi}), and the operator $\widehat{A}(\hat x,\hat y)$ reduces to (\ref{A-tilde-psi-f}) that implements the generalized hypergeometric differential equation (\ref{tilde-A-psi}). 

Consider now a symmetric quiver, whose structure is encoded in a symmetric matrix $C$. The motivic generating function associated to such a quiver takes form (\ref{P-C})
\begin{equation}
P_C(x_1,\ldots,x_m)=\sum_{d_1,\ldots,d_m} \frac{(-q^{1/2})^{\sum_{i,j=1}^m C_{i,j}d_id_j}}{(q;q)_{d_1}\cdots(q;q)_{d_m}} x_1^{d_1}\cdots x_m^{d_m},  \label{P-C-intro}
\end{equation}
and motivic Donaldson-Thomas invariants $\Omega_{d_1,\ldots,d_m;j}$ arise from the factorization of this series into a product of quantum dilogarithms (\ref{PQx-Omega}) \cite{Kontsevich:2010px,efimov2012}. Our main statement (\ref{psi-f-quiver}) is that for an arbitrary strip geometry, the brane partition function (\ref{psi-f-intro}) can be written in the form (\ref{P-C-intro}), with $x_1=q^{-(f+1)/2} x$ and $x_2,\ldots,x_m$ identified with $\alpha_i$ or $\beta_j$
\begin{equation}
\psi_f(x) = P_C(q^{-(f+1)/2}x, q^{-1/2}\alpha_1, \alpha_1, \dots, q^{-1/2}\alpha_r, \alpha_r, q^{-1/2}\beta_1, \beta_1, \dots, q^{-1/2}\beta_s, \beta_s), 
\end{equation}
for a particular choice of the quiver of size $1+2(r+s)$ defined by the matrix (\ref{C-psi-f}). This statement has deep consequences. In particular, Ooguri-Vafa BPS invariants for such a brane, which are also defined by the product decomposition into quantum dilogarithm, can be expressed in terms of combinations of motivic Donaldson-Thomas invariants associated to (\ref{P-C-intro}), and thus are immediately proven to be integer. It also follows that all generalized $q$-hypergeometric functions are determined by such motivic Donaldson-Thomas invariants. 

Having shown the above facts, we analyze various properties of brane partition functions and A-polynomials for quivers associated to strip geometries. This analysis is based on some general properties of A-polynomials and Donaldson-Thomas invariants for quivers, in particular the partial limit, which are interesting in their own right, and which we derive in section \ref{sec-quivers}. As one important outcome of this analysis we find a general expression for classical open BPS invariants in arbitrary framing $f$, for an arbitrary strip geometry with moduli $\alpha_1,\ldots,\alpha_r,\beta_1,\ldots,\beta_s$. These invariants are encoded in the product representation of the series $y=y(x)$ that is a solution of the classical mirror curve equation $A(x,y)=0$ in (\ref{A-hat-psi-f-class}), which arises in $q \to 1$ limit of (\ref{A-hat-intro}). This solution arises also from the limit of the ratio of brane partition functions (\ref{y-x-limit}) and it takes form (\ref{y-x-alpha-beta})
\begin{equation}
  y(x) = \lim_{q\to 1} \frac{\psi_f(qx)}{\psi_f(x)} =  \prod_{(n, l_1,\ldots,l_r, k_1,\ldots,k_s)>0} \left(1 - x^n \alpha_1^{l_1} \cdots \alpha_r^{l_r} \beta_1^{k_1}\cdots \beta_s^{k_s} \right)^{n\Omega_{n,l_1,\ldots,l_r, k_1,\ldots,k_s}},  
\end{equation}
and we show that open BPS invariants read (\ref{Omega-nlk-classical}) 
\begin{align}
\begin{split}
  \Omega_{n, l_1,\ldots,l_r, k_1,\ldots,k_s}  &= - \frac{1}{n} \sum_{i|{\rm gcd}(n, l_1,\ldots,l_r, k_1,\ldots,k_s)}\frac{(-1)^{f n/i} \mu(i)  }{(f+1)n + |l| + |k|}\binom{\left((f+1)n + |l| + |k|\right)/i}{n/i} \times \\
  &\quad \times\prod_{j=1}^r (-1)^{l_j/i}\binom{n/i}{l_j/i} \prod_{j=1}^s \frac{n}{n+k_j}\binom{(n+k_j)/i}{k_j/i},      \label{Omega-intro}
\end{split}
\end{align}
where $\mu(i)$ is the M{\"o}bius function, $|l|=\sum_i l_i$, and indices $n, l_1,\ldots,l_r, k_1,\ldots,k_s$ are associated to moduli $x,\alpha_1,\ldots,\alpha_r,\beta_1,\ldots,\beta_s$. The relation to quivers, and independently string theoretic interpretation, imply that (\ref{Omega-intro}) are integer, and therefore this expression provides a large set of number theoretic integrality statements: despite the factor of $1/n$ and other denominators, for each fixed $(r,s,f,n,l_1,\ldots,l_r, k_1,\ldots,k_s)$, the above expression must be integer. This vastly generalizes analogous statements for the framed unknot, or equivalently a brane in $\mathbb{C}^3$ or resolved conifold, presented in \cite{Garoufalidis:2015ewa,Basor:2017qxy,Luo:2016oza}.

\subsection{...a brief discussion...}

Let us also list a few questions that are motivated by our results, and which deserve further investigation. First, it should be understood in more detail how the structure of various objects assigned to quivers, e.g. moduli spaces of their representations or cohomologial Hall algebras, relates to topological string theory and properties of toric manifolds. Second, while in this paper we identify quivers corresponding to strip geometries, it is important to understand if analogous quiver description, or some generalization thereof, can be given for more general toric manifolds that contain compact four-cycles, such as the local $\mathbb{P}^2$, local $\mathbb{P}^1\times\mathbb{P}^1$, or local Hirzebruch surfaces. Third, it would be gratifying to provide more direct physical derivation of supersymmetric quantum mechanics associated to quivers that correspond to strip geometries, as well as more general topological string amplitudes, such as those that arise in the knots-quivers correspondence. Fourth, the role and the meaning of quivers that we identify should be understood in all other systems related to or engineered by topological string theory, such as supersymmetric gauge theories, vortex counting, etc. Fifth, it is of interest to understand if there are relations between quivers that we identify in this paper, and other quivers identified in related contexts \cite{Alim:2011ae,Manschot:2013sya,Eager:2016yxd,Zhu:2017lsn}. Sixth, it is tempting to relate the combinatorics of quivers that we identify to crystal models related to the topological vertex. Seventh, all these relations can be generalized to the refined case.

\subsection{...and a brief plan}

The plan of this paper is as follows. In section \ref{sec-top} we recall basics of topological string theory and properties of strip geometries, and then we compute brane amplitudes in such geometries, find corresponding quantum mirror curves, and discuss their limits. In section \ref{sec-quivers} we recall basics of quiver representation theory for symmetric quivers, introduce the partial classical limit, and assign quantum curves and A-polynomials to quivers. In section \ref{sec-top-quivers} we show that partition functions for branes in strip geometries can be expressed as motivic generating functions for quivers, and we identify the corresponding quivers. We also discuss general properties of quantum and classical mirror curves for strip geometries, and properties of BPS invariants, which follow from the relation to quiver representation theory. Finally, in section \ref{sec-examples} we consider several examples of strip geometries, and illustrate in such examples various structures introduced earlier. In appendix \ref{sec-app} we discuss various conventions related to the definition of motivic Donaldson-Thomas invariants and positivity of these invariants.

\section{Topological string theory, strip geometries, and brane amplitudes}   \label{sec-top}

Topological string amplitudes count, in appropriate sense, maps from Riemann surfaces into a target space. Open topological string amplitudes count maps from Riemann surfaces with boundaries, and the boundary conditions may be encoded by appropriately chosen branes. In this paper we consider A-model (holomorphic) amplitudes for target spaces which are toric Calabi-Yau threefolds that do not contain compact four-cycles; such manifolds are referred to as strip geometries or generalized conifolds. In this section we first briefly summarize the general structure of A-model amplitudes, as well as the topological vertex formalism and its simplifications that arise for strip geometries. We then compute the open partition function for the Aganagic-Vafa lagrangian brane in arbitrary framing and in arbitrary strip geometry. This result is given in (\ref{psi-hyper}) and it will be of our main interest in what follows. We also determine the quantum mirror curve operator (\ref{A-hat-psi-f}) that annihilates this brane partition function, identify the mirror curve that arises in the classical limit of this operator (\ref{A-hat-psi-f-class}), and find the differential operator (\ref{A-tilde-psi-f}) that arises in the modified classical limit. We also discuss the relation of partition functions (\ref{psi-hyper}) and the equations they satisfy to generalized hypergeometric functions and hypergeometric equations.

%***************************************************************************************************

\subsection{Topological string amplitudes and BPS invariants}

A-model topological string amplitudes depend on K{\"a}hler parameters $Q=\{Q_k\}$ of a given target Calabi-Yau manifold $M$, and open moduli $x=\{x_i\}$ that characterize branes. They are defined in terms of the genus expansion in the topological string coupling $\hbar$, and various terms in such expansion encode closed or open Gromov-Witten invariants. The full topological string amplitudes factorize into closed string contributions and -- in presence of branes -- open contributions, that involve both open and closed moduli 
\be
Z = Z^{\textrm{closed}}(Q) \cdot \psi^{\textrm{open}}(Q,x).   \label{Zclosedopen}
\ee
From the spacetime interpretation of topological strings \cite{Gopakumar:1998ii,Gopakumar:1998jq} it follows that topological string amplitudes can be expressed in a product form that represents counting of BPS states, in terms of the variable $q=e^{\hbar}$. In particular closed string contributions take form 
\be
Z^{\textrm{closed}}(Q) = \prod_{\beta\in H_2(M)}  \prod_j\prod_{l=1}^{\infty} (1 - Q^{\beta} q^{l+j})^{l N_{\beta,j}},    \label{Z-closed-GV}
\ee
where $N^m_{\beta}$ are conjecturally integer Gopakumar-Vafa invariants that count BPS states of closed M2-branes. Note that for fixed $\beta$ and $m$, the contribution from the product over $l$ is a generalization of the MacMahon function $M(q)=\prod_{l=1}^{\infty}(1-q^l)^l$ that counts plane partitions.

It is known that in certain systems open partition functions satisfy Schr{\"o}dinger-like equations, hence they are also referred to as wave-functions, and for this reason we denote them by the symbol $\psi^{\textrm{open}}(Q,x)$. Spacetime interpretation of BPS counting implies that in presence of branes open topological string amplitudes also have product decomposition. First, as argued in \cite{OoguriV,Labastida:2000zp,Labastida:2000yw}, the open partition function can be written in the form
\be
\psi^{\textrm{open}}(Q,x) = \sum_P \psi^{\textrm{open}}_P \textrm{Tr}_P X = \exp \Big(  \sum_{n=1}^\infty \sum_P \frac{1}{n} f_{P}(Q^n,q^n) \textrm{Tr}_P X^n  \Big),   \label{ZUV}
\ee
where we encoded brane moduli $x=\{x_i\}$ in a matrix $X=\textrm{diag}(x_1,x_2,\ldots)$, Young diagrams $P$ under summations represent brane boundary conditions, and
\be
f_{P}(Q,q) = \sum_{\beta,j} \frac{N_{P,\beta,j} Q^{\beta} q^j}{q^{1/2}-q^{-1/2}}  \label{fP}
\ee
are functions that encode integer multiplicities $N_{P,\beta,j}$ of open M2-branes in a relative class $\beta$, with spacetime spin $j$, and labeled by $P$. Multiplicities $N_{P,\beta,j}$ are referred to as Ooguri-Vafa invariants and they provide an interesting reformulation of open Gromov-Witten invariants. In the context of knots $N_{P,\beta,j}$ are also referred to as  Labastida-Mari{\~n}o-Ooguri-Vafa (LMOV) invariants, and $\psi^{\textrm{open}}_P$ are related to colored HOMFLY-PT polynomials \cite{OoguriV,Labastida:2000zp,Labastida:2000yw}. Taking advantage of the relation
\be
\textrm{Tr}_P X^n = \sum_{k^P} m_{k^P} \prod_i x_i^{n k^P_i},
\ee
where $k^P=\{k^P_i\}$ and $m_{k^P}$ are respectively weights of the representation $P$ and their multiplicities, the open partition function (\ref{ZUV}) can be written in the product form
\be
\psi^{\textrm{open}}(Q,x) = \prod_{P,j,\beta,k^P} \prod_{l=1}^{\infty} (1 - x^{k^P} Q^{\beta} q^{l+j-1/2}  )^{m_{k^P} N_{P,\beta,j}}.   \label{ZUV-prod}
\ee
Note that for fixed $P,j,\beta,k^R$, the product over $l$ represents the quantum dilogarithm (with appropriate arguments), which can also be written as a special case (with $n=\infty$) of the $q$-Pochhammer symbol
\be
(Q;q)_n = \prod_{i=0}^{n-1}(1-Qq^i).   \label{qPochhammer}
\ee

More precisely, a single trace $\textrm{Tr}_P X$ in (\ref{ZUV}) represents one stack of branes; for multiple stacks the open amplitude would in general take form
\be
\psi^{\textrm{open}}(Q,x) = \sum_{\{P_i\}} \psi^{\textrm{open}}_{\{P_i\}} \prod_i\textrm{Tr}_{P_i} X_i.
\ee
It is also convenient to write the total amplitude (\ref{Zclosedopen}), including both closed and open contributions, in the form
\be
Z = \sum_{\{P_i\}} Z_{\{P_i\}} \prod_i\textrm{Tr}_{P_i} X_i, \qquad Z_{\{P_i\}} = Z^{\textrm{closed}}\cdot \psi^{\textrm{open}}_{\{P_i\}}   \label{ZUV-2}
\ee

In this paper we consider mainly systems with a single brane. In this case $x$ is just a single variable. Then $\textrm{Tr}_P x \neq 0$ only for symmetric representations $P=S^n$, and $\textrm{Tr}_{S^n}(x) = x^n$, so that
\be
\psi^{\textrm{open}}(Q,x) = \prod_{n\geq 1;\beta,j} \prod_{k=1}^{\infty} \Big(1 - x^n Q^{\beta} q^{j+k-1/2} \Big)^{N_{n,\beta,j}}.
\label{Z-open-x} 
\ee

%***************************************************************************************************

\subsection{Topological vertex and strip geometries}

The structure of toric Calabi-Yau threefolds can be encoded in planar diagrams with trivalent vertices. Each edge (``leg'') of such a diagram represents a specific locus along which one circle in the toric fiber degenerates. Each trivalent vertex represents one $\mathbb{C}^3$ patch, and the whole diagram encodes the way in which such patches are glued. Topological string amplitudes for such threefolds can be computed by means of the topological vertex $C_{PQR}(q)$, which is the basic building block that gets associated to one trivalent vertex \cite{AKMV}. The topological vertex is labeled by three Young diagrams $P$, $Q$, and $R$, which are assigned respectively to the three legs of the trivalent vertex and encode relevant boundary conditions; moreover the topological vertex amplitude depends on the variable $q=e^{\hbar}$ that encodes the topological string coupling $\hbar$. The topological vertex amplitude has interpretation in terms of a plane partition with arbitrary boundary conditions at infinity encoded by diagrams $P,Q$ and $R$, and it can be expressed in terms of skew Schur functions $s_{P/S}$ \cite{Okounkov:2003sp}
\begin{equation}
C_{P Q R} (q) = q^{\frac{1}{2}(\kappa_{Q}+\kappa_{R})} s_{Q^{T}}(q^{\rho})
\, \sum_{S} s_{P/S}(q^{ Q^{T}+\rho}) s_{R^{T}/ S}(q^{ Q+\rho}),
\label{vertex}
\end{equation}
where $Q^T$ denotes a transpose of $Q$, $q^{Q+\rho} \equiv (q^{Q_1-1/2},q^{Q_2-3/2},q^{Q_3-5/2},\ldots)$, and 
\be
\kappa_R =  |R|+\sum_i R_i(R_i -2i) = -\kappa_{R^T},\quad\qquad |R|=\sum_i R_i.
\ee 
One can also consider more general framed vertex, with framing specified for each leg by integers $f_i$ for $i=1,2,3$, whose amplitude reads
\begin{equation}
C^{f_1,f_2,f_3}_{PQR} = (-1)^{f_1 |P|+f_2|Q|+f_3|R|} q^{(f_1 \kappa_{P}+f_2 \kappa_{Q}+f_3 \kappa_{R})
/2} C_{PQR}. \label{framing}
\end{equation}
The total amplitude for a given toric manifold is obtained by gluing such vertex amplitudes. Gluing of two vertices along an edge amounts to the identification (up to a transposition) of Young diagrams assigned to the two legs being glued, and resummation over all possible such diagrams. The edge (``internal leg'') that arises from such a gluing operation represents topologically $\mathbb{P}^1$, that arises from a circle in the toric fiber that degenerates at two vertices in question. 

We also recall that mirror manifolds to toric threefolds take form of algebraic varieties defined by one equation in four-dimensional complex space
\be
uv = A(x,y),  \label{mirror-curve}
\ee
where $A(x,y)$ is a polynomial in $x,y\in\mathbb{C}^*$, and the locus $A(x,y)=0$ is a Riemann surface referred to as the mirror curve. Mirror B-model topological string amplitudes can be computed by means of the topological recursion for the mirror curve. Mirror curves can also be quantized into difference operators $\widehat{A}(\hat x,\hat y)$ that impose difference equations for brane amplitudes \cite{ADKMV,abmodel}. In the tropical limit, in which pairs of pants arising from a decomposition of the Riemann surface reduce to trivalent vertices, the mirror curve reduces to the toric diagram of the original toric manifold.

\begin{figure}[h]
\begin{center}
\includegraphics[width=0.8\textwidth]{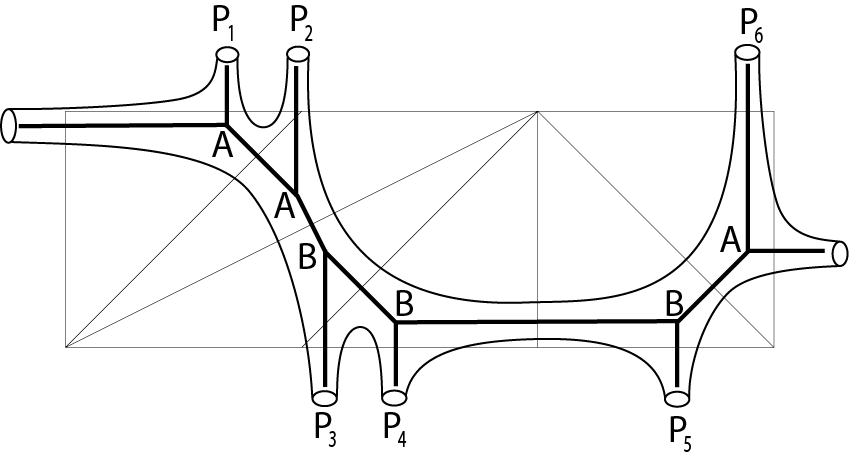} 
\caption{An example of a strip geometry. The toric diagram (made of thick segments) arises as the dual diagram to a triangulation of a rectangular strip (made of thin segments). Trivalent vertices in this case are respectively of type A, A, B, B, B, and A, and $\mathbb{P}^1$'s are represented by internal legs of type $(-2,0), (-1,-1), (-2,0),(-2,0),(-1,-1)$ (from left to right). Brane boundary conditions encoded in Young diagrams $P_i$ can be imposed at one external leg of each vertex. Thickening the toric diagram leads to a schematic picture of the mirror curve (shown in thin lines).}  \label{fig-strip}
\end{center}
\end{figure}

For toric threefolds that do not have compact four-cycles, toric diagrams take form of trees (without loops). As the legs of the diagram should not intersect, apart from the closed topological vertex geometry (which involves one vertex connected via three legs to three other vertices), all other such manifolds are necessarily so called strip geometries (also called generalized conifolds), whose toric diagrams arise as dual diagrams to a triangulation of a strip, as shown in fig. \ref{fig-strip}. A toric diagram for strip geometry consists of a chain of legs that represent various $\mathbb{P}^1$'s, which locally represent either the resolved conifold or the resolution of $\mathbb{C}^3/\mathbb{Z}_2$, and which are referred to respectively as $(-2,0)$ and $(-1,-1)$ curves. An example of a strip geometry and the corresponding mirror curve are shown in fig. \ref{fig-strip}. Topological vertex computations for such geometries can be partly conducted and simplified, as explained in \cite{Iqbal:2004ne}.

Let us recall how to compute the total (including open and closed contributions) topological string amplitude (\ref{ZUV-2}) for a strip geometry, following \cite{Iqbal:2004ne}. Each strip consists of a series of topological vertices. Each two neighboring vertices are connected by an internal leg that represents $\mathbb{P}^1$ of type $(-2,0)$ and $(-1,-1)$, with K{\"a}hler parameter $Q_k$. Apart from the two external vertices, two legs of each of the other (internal) vertices are connected to its immediate neighbors, while the third leg is external and can encode arbitrary brane boundary conditions. Therefore one might assume that brane boundary conditions for the $i$'th external leg of the $i$'th vertex (including two external vertices, for which we choose one particular external leg), are labeled by arbitrary Young diagram $P_i$. It then follows that the full amplitude is a product of several factors. First, each vertex contributes the Schur function $s_{P_i}(q^{\rho}) \equiv s_{P_i}(q^{-1/2},q^{-3/2},q^{-5/2},\ldots)$. Second, consider a pair of vertices from the strip with attached Young diagrams $P_i$ and $P_j$, and define
\be
\{ P_i P_j \} = \prod_k (1 - Q_{ij} q^k)^{C_k(P_i,P_j)} \exp\Big( \sum_{m=1}^{\infty} \frac{Q_{ij}^m}{m(2\sin \frac{m \hbar}{2})^2}  \Big) , \label{bracket}
\ee
where $Q_{ij}=Q_iQ_{i+1}\cdots Q_{j-1}$ is the product of K{\"a}hler parameters $Q_k$ associated to internal legs that join the pair of vertices under consideration, and the exponents $C_k(P,R)$ are defined by
\be
\sum_k C_k(P,R) q^k = \frac{q}{(q-1)^2} \Big(1 + (q-1)^2 \sum_{i=1}^{d_P} q^{-i}  \sum_{j=0}^{P_i-1} q^j   \Big)  \Big(1 + (q-1)^2 \sum_{i=1}^{d_R} q^{-i}  \sum_{j=0}^{R_i-1} q^j   \Big) - \frac{q}{(1-q)^2}   \label{Ck-define}
\ee
where $d_P$ denotes the number of rows in the Young diagram $P$. Furthermore, to the first vertex in a strip we assign a type A or B, if respectively its amplitude can be written in the form $C_{S\bullet P}$ or $C_{\bullet S P}$ (where diagrams $S$ are summed over in the internal leg, and $P$ labels an external leg). We also assign types A or B to all other vertices recursively: the next vertex has the same type as the preceding one if they are connected by $\mathbb{P}^1$ of type $(-2,0)$, and it is assigned an opposite type if two vertices are connected by $\mathbb{P}^1$ of type $(-1,-1)$. Then each pair of vertices with boundary conditions $P_i$ and $P_j$ contributes to the amplitude a factor, which depends on the types of these two vertices; for a pair of vertices of types $(A,A)$, $(A,B)$, $(B,A)$, $(B,B)$, this contribution respectively takes form $\{P_i,P_j^T\}^{-1}$, $\{P_i,P_j\}$, $\{P_i^T,P_j^T\}$, $\{P_i^T,P_j\}^{-1}$, where $P^T$ denotes a transposition of a diagram $P$. 

To sum up, the total topological string amplitude (\ref{ZUV-2}) for a strip geometry, with boundary conditions at the $i$'th vertex encoded in a Young diagram $P_i$, takes form
\be   
Z_{\{P_i\}} = \prod_i s_{P_i}(q^{\rho}) \prod_{i,j} \{P_i^*,P_j^*\}^{\pm 1},   \label{Z-strip}
\ee
where powers $\pm 1$, as well as $P^*$ that denote either just $P$ or $P^T$, depend on types ($A$ or $B$) of vertices $i$ and $j$. Note that this result involves both open and closed contributions, and the latter ones arise only from the exponential factors in (\ref{bracket}) and can be rewritten in the product form (\ref{Z-closed-GV}). As an example, the partition function for the toric manifold in fig. \ref{fig-strip} reads
\be
Z_{P_1,\ldots,P_6} = \frac{\{P_1 P_3\} \{P_1 P_4\} \{P_1 P_5\} \{P_2 P_3\} \{P_2 P_4\} \{P_2 P_5\}   \{P_3^T P_6^T\} \{P_4^T P_6^T\} \{P_5^T P_6^T\}}{\{P_1 P_2^T\} \{P_1P_6^T\} \{P_2P_6^T\} \{P_3^T P_4\} \{P_3^T P_5\} \{P_4^T P_5\}} \prod_{i=1}^{6} s_{P_i}(q^{\rho})  .
\ee

%***************************************************************************************************

\subsection{Brane amplitudes and generalized $q$-hypergeometric functions}

We now focus on a particular amplitude we are interested in, which involves open contributions for one brane in arbitrary framing attached to the first vertex. Without loss of generality we assume that the first vertex is of type A, and it is labeled by a Young diagram $P$. We also assume that the strip consists in total of $1+r+s$ vertices, and apart from the first one of type A, there are $s$ other vertices of type A and $r$ vertices of type B. We denote open contributions to the amplitude (\ref{Z-strip}) by $\psi_P^{\textrm{open}}$, and they are obtained simply by removing all exponential factors that arise from (\ref{bracket}) from the resulting total amplitude. Furthermore, we are interested only in the single framed brane generating function that is defined as a resummation with a single generating parameter, which for convenience we denote $q^{-1/2}x$. 

\begin{figure}[h]
\begin{center}
\includegraphics[width=0.8\textwidth]{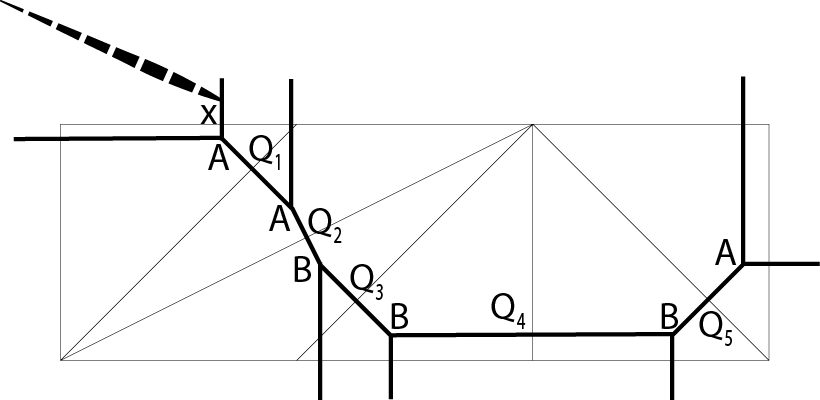} 
\caption{A strip geometry with a single brane at the first vertex. The brane modulus is denoted by $x$, and internal segments represent $\mathbb{P}^1$'s with K{\"a}hler parameters $Q_k$.}  \label{fig-strip-2}
\end{center}
\end{figure}

Taking into account the framing factor (\ref{framing}) and denoting framing by $f\in\mathbb{Z}$, such a generating function takes form
\be
\psi_f(x) = \sum_P (-1)^{f |P|} q^{f \kappa_{P}/2} s_P(q^{-1/2}x) \psi_P^{\textrm{open}} = \sum_n (-1)^{fn} q^{fn(n-1)/2} (q^{-1/2} x)^n \psi_{(n)}^{\textrm{open}},
\ee
where we used the fact that $s_P(x)=x^n$ when $P=(n)$ consists of only one row of length $n$, and for other Young diagrams $s_P(x)$ is zero. The factor $\psi_{(n)}^{\textrm{open}}$ above therefore denotes the amplitude with a single brane in the trivial framing, at the first vertex, labeled by a Young diagram with one row of length $n$, and its explicit form arises from the following specialization of (\ref{Z-strip}). First, it involves only one Schur function 
\be
s_{(n)}(q^{\rho}) = \frac{(-1)^n q^{n^2/2}}{(q;q)_n},
\ee
where $(q;q)_n = \prod_{k=1}^n (1-q^k)$ is a special case of the $q$-Pochhammer symbol defined in (\ref{qPochhammer}). Second, in this case all factors $\{P_i^*,P_j^*\}^{\pm 1}$ take form either $\{\bullet,\bullet \}^{\pm 1}$ (with the argument $Q_{ij}$ and with $\bullet$ denoting the empty partition) if $i,j\neq 1$ (i.e. the first vertex is not involved), or $\{(r),\bullet\}^{\pm 1}$ (with the argument $Q_{1j}$) if the pair involves the first and the $j$'th vertex in the strip. In the former case all $C_k(\bullet,\bullet)=0$, so that $\{\bullet,\bullet \}^{\pm 1}$ reduces to the closed string contribution that we ignore in the computation of $\psi_{(n)}^{\textrm{open}}$. In the latter case the coefficients (\ref{Ck-define}) take form
\be
\sum_k C_k\big((r),\bullet\big) q^k = \frac{1-q^n}{1-q} = 1 + q + \ldots + q^{n-1},
\ee
i.e. $C_k\big((n),\bullet\big)=1$ for $0\leq k<n$, and $C_k\big((n),\bullet\big)=0$ for $k\geq n$, and in such case
\be
\{(n),\bullet\} = \prod_{k=0}^{n-1} (1 - Q_{1j} q^k) \exp\Big( \sum_{m=1}^{\infty} \frac{Q_{1j}^m}{m(2\sin \frac{m \hbar}{2})^2}  \Big) \equiv (Q_{1j};q)_n \exp\Big( \sum_{m=1}^{\infty} \frac{Q_{1j}^m}{m(2\sin \frac{m \hbar}{2})^2}  \Big)  ,
\ee
so that the contribution to the open amplitude is simply given by the $q$-Pochhammer $(Q_{1j};q)_n$ in appropriate power $\pm 1$. As we assumed that the first vertex is of type A, such $q$-Pochhammer factors arise in power $\pm 1$, respectively if the $j$'th vertex is of type B or A. For simplicity we also denote by $\alpha_i$, for $i=1,\ldots,r$, all $Q_{1j}$ for which the $j$'th vertex is of type B, and by $\beta_i$, for $i=1,\ldots,s$, those $Q_{1j}$ for which the $j$'th vertex is of type A. With this notation, and taking into account all factors discussed above, the framed brane generating function takes form
\begin{align}
\begin{split}
\psi_f(x) &= \sum_{n=0}^{\infty} \big((-1)^{n} q^{n(n-1)/2}  \big)^{f+1}  \frac{ x ^n}{(q;q)_n} \prod_{j} (Q_{1j};q)^{\pm 1}_n  = \\
&= \sum_{n=0}^{\infty} \big((-1)^{n} q^{n(n-1)/2}  \big)^{f+1}  \frac{ x ^n}{(q;q)_n} \frac{(\alpha_1;q)_n (\alpha_2;q)_n\cdots (\alpha_r;q)_n}{(\beta_1;q)_n (\beta_2;q)_n\cdots (\beta_s;q)_n}. \label{psi-hyper}
\end{split}
\end{align}

This is a very interesting result on which our analysis in what follows will be based. Notice that there may exist several strip geometries -- which are related by flop transitions -- for which the brane amplitude takes the same form given in the second line above. Nonetheless, brane partition functions for such geometries differ in a way in which K{\"a}hler parameters $Q_k$ are related to $\alpha_i$ and $\beta_j$. We present examples of such geometries in section \ref{sec-examples}.

Furthermore, note that for appropriate choice of framing the result (\ref{psi-hyper}) reduces to the generalized $q$-hypergeometric function. The most common definition of such a function  \cite{gasper2004basic} arises for $f=s-r$
\begin{align}
\begin{split}
\psi_{s-r}(x) &= \pphiq{r}{s}{\alpha_1,\alpha_2, \dots, \alpha_r}{\beta_1, \beta_2, \dots, \beta_r}{q}{x}  = \\
&= \sum_{n=0}^{\infty} \big((-1)^{n} q^{n(n-1)/2}  \big)^{1+s-r}  \frac{ x ^n}{(q;q)_n} \frac{(\alpha_1;q)_n (\alpha_2;q)_n\cdots (\alpha_r;q)_n}{(\beta_1;q)_n (\beta_2;q)_n\cdots (\beta_s;q)_n},    \label{q-hyper}
\end{split}
\end{align}
and this is a definition of $q$-hypergeometric functions we will refer to in what follows (note that sometimes these functions are defined without including the factor $(-1)^{n} q^{n(n-1)/2}$, which in our convention amounts to setting framing to $f=-1$). 

For example, in fig. \ref{fig-strip-2} we have $r=3, s=2$, and so the generating function for a single brane in framing $f=s-r=-1$  takes form
\be
\psi_{-1}(x) = \pphiq{3}{2}{\alpha_1,\alpha_2, \alpha_3}{\beta_1, \beta_2}{q}{x},
\ee
where $\alpha_1 = Q_1 Q_2$, $\alpha_2 = Q_1 Q_2 Q_3$, $\alpha_3 = Q_1Q_2Q_3Q_4$, and $ \beta_1 = Q_1$, $\beta_2 = Q_1 Q_2 Q_3 Q_4 Q_5$.

%***************************************************************************************************

\subsection{Quantum mirror curves and generalized hypergeometric equations} 

Once we have derived the brane partition function (\ref{psi-hyper}), we can also find a $q$-difference equations it satisfies. Such $q$-difference equations are interpreted as quantum mirror curves, and in the $q\to 1$ limit they should reduce to (classical) mirror curves \cite{ADKMV,abmodel}. For strip geometries we can identify such curves explicitly. To this end we write $\psi_f(x) = \sum_n p_n x^n$, where $p_n$ is identified with the summand (without $x^n$ factor) in (\ref{psi-hyper}), and we note that $p_n$ satisfies the relation
\be
p_{n+1} (1-q^{n+1})\prod_{j=1}^s (1- \beta_j q^n) = p_n (-1)^{f+1} q^{n(f+1)} \prod_{j=1}^r (1-\alpha_j q^n).
\ee
Multiplying both sides of this relation by $x^{n+1}$, summing over all $n$, and recalling that $\hat y f(x)=f(qx)$, we find the operator
\be
\widehat{A}(\hat x,\hat y) = (1 - \hat y)\prod_{j=1}^s (1 - q^{-1} \beta_j \hat y) + (-1)^{f} \hat{x}\,  \Big(\prod_{j=1}^r (1-\alpha_j\hat y) \Big) \hat{y}^{f+1}  \label{A-hat-psi-f}
\ee
that annihilates the brane partition function (\ref{psi-hyper}) 
\be
\widehat{A}(\hat x,\hat y) \psi_f(x) = 0.
\ee
We refer to (\ref{A-hat-psi-f}) as the quantum mirror curve. Note that for $f=s-r$ it reduces to the operator that imposes the generalized $q$-hypergeometric equation for the $q$-hypergeometric function (\ref{q-hyper}) \cite{gasper2004basic}. 

Clearly, and as expected, for $q\to 1$ the operator $\widehat{A}(\hat x,\hat y)$ reduces to the mirror curve for a given strip geometry
\be
A(x,y) = (1-y)\prod_{j=1}^s (1 - \beta_j  y) + (-1)^{f} x y^{f+1} \prod_{j=1}^r (1-\alpha_j y) = 0.    \label{A-hat-psi-f-class}
\ee
Solving this equation for $y=y(x)$ we obtain a function which can be thought of as the classical limit of the operator $\hat{y}$, and it can also be obtained as the appropriate ratio of brane partition functions (\ref{psi-hyper})
\be
y(x)=\lim_{q\to 1} \frac{\hat{y}\psi_f(x) }{\psi_f(x)} = \lim_{q\to 1}\frac{\psi_f(qx) }{\psi_f(x)}.  \label{y-x-limit}
\ee
Taking advantage of the relation to quivers, we will find an explicit expression for coefficients of the series $y(x)$ in (\ref{y-c_i}).

Furthermore, apart from the classical limits $q\to 1$ in which all other parameters are kept fixed, it is also of interest to consider a limit in which $q$-difference equations reduce to differential equations. In this limit we also take $q=e^{\hbar}\to 1$, however first we appropriately rescale various variables and parameters. Considering two terms in (\ref{A-hat-psi-f}), we find that after setting
\be
  x \to (q-1)^{1+s-r} x, \quad \alpha_i= q^{a_i}, \quad \beta_j = q^{b_j}, \label{rescale}
\ee
writing $\hat{y}=e^{\hbar x\partial_x}$, and expanding in $\hbar=\log q$, the leading term in $\hbar$ expansion (note that the rescaling of $x$ is crucial in getting this result) reduces to a non-trivial differential operator
\be
\widetilde{A} =  \partial_x  \prod_{j=1}^s (x\partial_x +b_j - 1) + (-1)^{1+s-r-f}  \prod_{j=1}^r (x\partial_x + a_j ) . \label{A-tilde-psi-f}
\ee
This operator imposes the differential equation
\be
\widetilde{A} \widetilde{\psi}(x)  = 0  \label{tilde-A-psi}
\ee
for the function that arises as the limit (\ref{rescale}) of (\ref{psi-hyper})
\be
\widetilde{\psi}(x) = \sum_{n=0}^{\infty} (-1)^{n(s-r-f)}\frac{x^n}{n!} \frac{(a_1)_n (a_2)_n \dots (a_r)_n}{(b_1)_n \dots (b_s)_n},   \label{tilde-psi}
\ee
where we used that $(q^a,q)_n\simeq (-\hbar)^n(a)_n$, and $(a)_n=\prod_{i=0}^{n-1} (a+i)$ is the ordinary Pochhammer symbol. Note that the operator (\ref{A-tilde-psi-f}) and the function (\ref{tilde-psi}) depend on $f$ in a very minor way. In particular for $f=s-r$ the function $\widetilde{\psi}(x)$ reduces to the generalized hypergeometric function $_rF_s$, which we obtain as the limit of the generalized $q$-hypergeometric function (\ref{q-hyper})
\begin{align}
\begin{split}
&  \lim_{q\rightarrow 1} \pphiq{r}{s}{q^{a_1}, q^{a_2}, \dots, q^{a_r}}{q^{b_1}, \dots, q^{b_s}}{q}{(q-1)^{1+s-r} x} =  \\
& \qquad = \pFq{r}{s}{a_1, a_2, \dots, a_r}{b_1, \dots, b_s}{x} 
 = \sum_{n=0}^{\infty} \frac{x^n}{n!} \frac{(a_1)_n (a_2)_n \dots (a_r)_n}{(b_1)_n \dots (b_s)_n}. \label{Frs}
\end{split}
\end{align}
The equation (\ref{tilde-A-psi}) for $f=s-r$ is nothing but the generalized hypergeometric equation.

%***************************************************************************************************
%***************************************************************************************************

\section{Quivers, Donaldson-Thomas invariants, and A-polynomials}    \label{sec-quivers}

We now summarize some aspects of a seemingly unrelated theory of quiver representations \cite{Kontsevich:2010px,COM:8276935,Rei12}. One of the aims of this theory is to characterize properties of the moduli space of representations of a given quiver. Such properties -- in particular homological structure of the moduli space -- are encoded in motivic Donaldson-Thomas invariants, which can be explicitly determined in particular for a large class of symmetric quivers. Apparently, such symmetric quivers arise in connection with brane amplitudes, as has been shown in the knots-quivers correspondence in \cite{Kucharski:2017poe,Kucharski:2017ogk}, and as we discuss in what follows in more general context of topological string theory. 

After reviewing basic features of representations of symmetric quivers and their Donaldson-Thomas invariants in section \ref{ssec-DT}, in section \ref{ssec-partial} we introduce a novel limit that we refer to as the partial classical limit. We will show in section \ref{sec-top-quivers} that this partial limit enables to determine explicitly a solution of the mirror curve equation and classical Ooguri-Vafa invariants for an arbitrary strip geometry. 

Furthermore, in section \ref{ssec-A-quivers} we show that certain specializations of quiver generating series satisfy difference equations that can be interpreted as quantum curves, and which reduce to differential or algebraic equations in appropriate limits. We refer to operators that implement these equations as quantum or classical A-polynomials for quivers.

%***************************************************************************************************

\subsection{Motivic and numerical Donaldson-Thomas invariants for quivers}  \label{ssec-DT}

Let us focus on symmetric quivers with $m$ vertices, whose structure we encode in a symmetric square matrix $C$ of size $m$ with integer entries. The element $C_{i,j}$ of this matrix denotes the number of arrows from vertex $i$ to vertex $j$. To this quiver one associates a motivic generating series, defined by
\begin{equation}
P_C(x_1,\ldots,x_m)=\sum_{d_1,\ldots,d_m} \frac{(-q^{1/2})^{\sum_{i,j=1}^m C_{i,j}d_id_j}}{(q;q)_{d_1}\cdots(q;q)_{d_m}} x_1^{d_1}\cdots x_m^{d_m}.   \label{P-C}
\end{equation}
This generating function has a product decomposition
\be
P_C(x_1,\ldots,x_m)=
\prod_{(d_1,\ldots,d_m)\neq 0} \prod_{j\in\mathbb{Z}} \prod_{k=1}^{\infty} \Big(1 -  \big( x_1^{d_1}\cdots x_m^{d_m} \big) q^{k+(j-1)/2} \Big)^{(-1)^{j+1}\Omega_{d_1,\ldots,d_m;j}},   \label{PQx-Omega}
\ee
which defines motivic Donaldson-Thomas invariants. More precisely, motivic Donaldson-Thomas invariants are simple redefinitions of $\Omega_{d_1,\ldots,d_m;j}$ introduced via the above decomposition, as we discuss in detail in appendix \ref{sec-app}; however for brevity we also refer to $\Omega_{d_1,\ldots,d_m;j}$ simply as motivic Donaldson-Thomas invariants. It is conjectured in \cite{Kontsevich:2010px} and proven in \cite{efimov2012} that motivic Donaldson-Thomas invariants (identified in appendix \ref{sec-app}), or equivalently combinations $(-1)^{d_1+\ldots+d_m}\Omega_{d_1,\ldots,d_m;j}$, are positive integers. Motivic Donaldson-Thomas invariants $\Omega_{d_1,\ldots,d_m;j}$ of a symmetric quiver can be interpreted as the intersection Betti numbers of the moduli space of its semisimple representations, or as the Chow-Betti numbers of the moduli space of all simple representations \cite{MR,FR}. Interestingly, quiver generating functions (\ref{P-C}) take form of generalized Nahm sums \cite{Nahm}, which may indicate their relations to other systems in which such sums arise.

In the classical limit $q\rightarrow 1$ motivic Donaldson-Thomas invariants reduce to numerical Donaldson-Thomas invariants, which are encoded in the classical generating series defined by the ratio
\begin{equation}
y(x_1, \dots, x_m) = \lim_{q\to 1} \frac{P_C(q x_1, \dots, q x_m)}{P_C(x_1, \dots, x_m)} \equiv \sum_{l_1, \dots, l_m} b_{l_1, \dots, l_m} x_1^{l_1}\cdots x_m^{l_m}.   \label{y-x1-xm}
\end{equation}
In what follows we refer to this limit as the \emph{complete} classical limit. Numerical Donaldson-Thomas invariants $\Omega_{d_1, \dots, d_m}$ are then encoded in the following product decomposition of the above classical generating series
\begin{equation}
  y(x_1, \dots, x_m)  = \prod_{(d_1, \dots, d_m)\neq 0} \left(1- (x_1^{d_1}\cdots x_m^{d_m})\right)^{(d_1 + \dots + d_m)\Omega_{d_1, \dots, d_m}}. \label{y-class-product}
\end{equation}
Note that numerical Donaldson-Thomas invariants $\Omega_{d_1, \dots, d_m}$ are combinations of the motivic ones
\begin{equation}
  \Omega_{d_1, \dots, d_m} = \sum_j (-1)^j \Omega_{d_1, \dots, d_m;j}.   \label{Omega-numerical}
\end{equation}

In \cite{Panfil:2018sis} explicit expressions for coefficients $b_{l_1, \dots, l_m}$ in (\ref{y-x1-xm}) and classical invariants $\Omega_{d_1, \dots, d_m}$ for an arbitrary symmetric quiver have been found. The former ones take form
\begin{equation}
b_{l_1, \dots, l_m} = A(l_1, \dots, l_m) \prod_{j=1}^m \frac{(-1)^{(C_{j,j}+1)l_j}}{1 + \sum_{i=1}^m C_{i,j} l_i}\binom{1 + \sum_{i=1}^m C_{i,j}}{l_j},   \label{b-l1-lm}
\end{equation}
where $A(l_1, \dots, l_m)\equiv A_C(l_1, \dots, l_m) $ are polynomials of degree $m-1$ whose coefficients depend on entries of the matrix $C$, and which are defined inductively by
\begin{equation}
    A_C(l_1, \dots, l_{m-1}, 0) = A_{C'}(l_1, \dots, l_{m-1}) \Big(1 + \sum_{i=1}^{m-1} C_{i,m} l_i \Big),
\end{equation}
where $C'$ is the submatrix of $C$ made of its first $m-1$ rows and columns, and
with the initial condition $A(l_1) = 1$. These polynomials are defined uniquely once their invariance under permutations $\sigma\in S_m$ is imposed, $A_{\sigma\circ C}(x_{\sigma_1,\ldots,x_{\sigma_m}}) = A_C(x_1,\ldots,x_m)$, where $[\sigma\circ C]_{i,j}=C_{\sigma_i,\sigma_j}$.
It also follows that $y(x_1, \dots, x_m)\equiv y_C(x_1, \dots, x_m)$ are invariant under the action of $\sigma\in S_m$
\be
y_{\sigma\circ C}(x_{\sigma_1},\ldots,x_{\sigma_m}) = y_C(x_1,\ldots,x_m).   \label{y-invariance}
\ee

For example, for a quiver with one vertex and $\alpha$ loops, encoded in the matrix $C=[\alpha]$, we get
\begin{equation}
b_i = \frac{(-1)^{(\alpha+1) i}}{\alpha i + 1}\binom{\alpha i + 1}{i},   \label{b-i}
\end{equation}
and for a symmetric quiver with two vertices encoded in the matrix $C=\bigl[\begin{smallmatrix}
\alpha&\beta \\ \beta&\gamma
\end{smallmatrix} \bigr]$ we find
\begin{equation}
  b_{i,j} = \frac{(-1)^{(\alpha+1)i + (\gamma+1)j}  (\beta i + \beta j + 1)} {(\alpha i + \beta j + 1)(\beta i + \gamma j + 1)}\binom{\alpha i + \beta j + 1}{i}\binom{\beta i + \gamma j + 1}{j}.    \label{bij}
\end{equation}
We write down explicit formulas for numerical Donaldson-Thomas invariants $\Omega_{d_1, \dots, d_m}$ in (\ref{Omega-d1-dm}).

%***************************************************************************************************

\subsection{Partial classical limit}    \label{ssec-partial}

In the classical limit that defines the classical generating function $y(x_1,\ldots,x_m)$ each variable $x_i$ is treated in the same way, and gets multiplied by $q$ in $P_C(x_1,\ldots,x_m)$ in the numerator in (\ref{y-x1-xm}). However in the context of topological string amplitudes we will consider quiver generating functions in which one variable plays a special role, and it is of interest to consider a limit in which only such variable gets multiplied by $q$. This motivates us to introduce the \emph{partial} classical limit of the quiver generating function 
\begin{equation}
  y_j(x_1, \dots, x_m) = \lim_{q\to 1} \frac{P_C(x_1, \dots, x_{j-1}, q x_j, x_{j+1}, \dots, x_m)}{P_C(x_1, \dots, x_m)} \equiv \sum_{l_1, \dots, l_m} c^{(j)}_{l_1, \dots, l_m} x_1^{l_1}\cdots x_m^{l_m} ,   \label{y-j}
\end{equation}
where in the numerator only $x_j$ is multiplied by $q$. A simple computation involving the product decomposition (\ref{PQx-Omega}) and then taking the classical limit shows that $y_j(x_1, \dots, x_m)$ has an analogous product decomposition to (\ref{y-class-product}) 
\begin{equation}
y_i(x_1, \dots, x_m) = \prod_{(d_1, \dots, d_m)\neq 0} \left(1- (x_1^{d_1}\cdots x_m^{d_m})\right)^{d_i \Omega_{d_1, \dots, d_m}},   \label{y_i-Omega}
\end{equation}
where $\Omega_{d_1, \dots, d_m}$ are the same numerical Donaldson-Thomas invariants as in (\ref{y-class-product}). From this decomposition we immediately deduce that (\ref{y-class-product}) is simply the product of $y_j(x_1, \dots, x_m)$
\begin{equation}
  y(x_1, \dots, x_m) = \prod_{j=1}^{m}  y_j(x_1, \dots, x_m).   \label{y-prod-yj}
\end{equation}
It is also easy to see that the following relations hold
\begin{align}
\begin{split}
 & \ \qquad y_j(x_1,\dots, x_{j-1}, 0, x_{j+1},\dots, x_n) = 1, \\
&   y_j(0,\dots, 0, x_j, 0, \dots, 0) = y(0,\dots, 0, x_j, 0, \dots, 0).   
\end{split}
\end{align}

The functions $y_j(x_1, \dots, x_m)\equiv y_{j;C}(x_1, \dots, x_n)$ depend on the matrix $C$ and for various $j$ they are related by symmetry operations. As above, consider a permutation $\sigma \in S_m$ that acts on matrices as $[\sigma \circ C]_{i,j} = C_{\sigma_i, \sigma_j}$. Partial classical limits are covariant under this symmetry operation
\begin{equation}
  y_{j;C}(x_1, \dots, x_m) = y_{\sigma_j;\sigma \circ C}(x_{\sigma_1}, \dots, x_{\sigma_m}),  \label{y-j-cov}
\end{equation}
so that all $y_j(x_1, \dots, x_m)$ are determined e.g. by $y_1(x_1, \dots, x_m)$. For example, for $m=2$ and $C=\bigl[\begin{smallmatrix}
\alpha&\beta \\ \beta&\gamma
\end{smallmatrix} \bigr]$ and $C'=\bigl[\begin{smallmatrix}
\gamma&\beta \\ \beta&\alpha
\end{smallmatrix} \bigr]$
we get
\begin{equation}
  y_{2;C}(x_1, x_2) = y_{1;C'}(x_2, x_1).    \label{y2-y1}
\end{equation}
The covariance of the partial limits under the action of the permutation group (\ref{y-j-cov}) implies the invariance of the complete classical limit (\ref{y-invariance}).

We now postulate explicit expressions for coefficients $c^{(j)}_{l_1, \dots, l_m}$ of functions (\ref{y-j}). In view of the symmetry properties discussed above, it is sufficient to determine $c^{(1)}_{l_1, \dots, l_m}$. We find that its form is similar to (\ref{b-l1-lm})
\begin{equation}
  c^{(1)}_{l_1, \dots, l_m} = A_{1;C}(l_1, \ldots, l_m)\prod_{j=1}^m \frac{(-1)^{(C_{j,j}+1) l_j}}{\delta_{1,j} + \sum_{i=1}^m C_{i,j}l_i} \binom{\delta_{1,j} + \sum_{i=1}^m C_{i,j}l_i}{l_j},   \label{c-1-l1-lm}
\end{equation}
where $A_{1;C}(l_1, \dots, l_m)$ is a homogeneous polynomial of degree $m-1$ in variables $l_i$. This polynomial is invariant under the action of a subset of permutations acting on all but the first variable, it satisfies the relation
  \begin{equation}
    A_{1;C}(0, l_2, \dots, l_m) = 0,
  \end{equation}
and it is defined inductively
  \begin{equation}
    A_{1;C}(l_1, \dots, l_{m-1}, 0) = A_{1;C'}(l_1, \dots, l_{m-1}) \sum_{i=1}^{m-1} C_{i, m} l_i,
  \end{equation}
where the matrix $C'$ arises from removing the last row and the last column from the matrix $C$. The initial condition for this recursion reads $A_{1;C}(l_1) = 1$. These conditions suffice to construct polynomials $A_{1;C}(l_1, \dots, l_m)$. Note also that
\begin{equation}
  c_{0,\dots, 0} = 1, \qquad c_{0, l_2, \dots, l_m} = 0 \ \,  \textrm{for}\ l_2, \ldots, l_m > 0.
\end{equation}

To sum up, we determined coefficients (\ref{c-1-l1-lm}) in the expansion of the $y_1(x_1,\ldots,x_m)$ function defined in (\ref{y-j}). We can also relate these coefficients to classical Donaldson-Thomas invariants (\ref{Omega-numerical}), which appear also in (\ref{y_i-Omega}). To this end it is useful to compute first the logarithm 
\begin{equation}
  \log y_1(x_1, \dots, x_m) = \sum_{(d_1, \dots, d_m)>0} d_{d_1, \dots, d_m}^{(1)} \prod_{j=1}^m x_j^{d_j}.  \label{log-y_1}
\end{equation}
The coefficients in this expression are closely related to those in (\ref{c-1-l1-lm}), analogously as discussed in \cite{Panfil:2018sis}, and we find that they take form
\begin{equation}
  d_{d_1, \dots, d_m}^{(1)} = A_{1;C}(d_1, \dots, d_m) \prod_{j=1}^m \frac{1}{ \sum_{i=1}^m C_{i,j} d_i}\binom{\sum_{i=1}^m C_{i,j} d_i}{d_j},    \label{log-y_1-d}
\end{equation}
where $A_{1;C}(d_1, \dots, d_m)$ is the same polynomial as in (\ref{c-1-l1-lm}). It then follows that the classical Donaldson-Thomas invariants take form
\begin{align}
\begin{split}
  \Omega_{d_1, \dots, d_m} &= \, \frac{1}{d_1} \sum_{i|{\rm gcd}(d_1, \dots, d_m)} \frac{\mu(i)}{i} d_{d_1/i,\dots, d_m/i}^{(1)} =\\
  &= - \frac{1}{d_1} \sum_{i|{\rm gcd}(d_1, \dots, d_m)} \mu(i)\,  A_{1;C}(d_1, \dots, d_m) \prod_{j=1}^m \frac{1}{ \sum_{i=1}^m C_{i,j} d_i}\binom{\sum_{i=1}^m C_{i,j} d_i/i}{d_j/i},   \label{Omega-d1-dm}
\end{split}
\end{align}
where $\mu(i)$ is the M{\"o}bius function, and we used the fact that $A_{1;C}(d_1, \dots, d_m)$ is a homogeneous polynomial of degree $m-1$, so that
\begin{equation}
  A_{1;C}(d_1/i, \dots, d_m/i) \prod_{j=1}^m \frac{1}{ \sum_{i=1}^m C_{i,j} d_i/i} = i\, A_{1;C}(d_1, \dots, d_m) \prod_{j=1}^m \frac{1}{ \sum_{i=1}^m C_{i,j} d_i}.
\end{equation}

Let us illustrate the above result for quivers of small size. For $m=1$ and $C=[\alpha]$ we get of course the same result as in (\ref{b-i})
\begin{equation}
  y_1(x) = y(x) = \sum_{i=0}^{\infty} \frac{(-1)^{(\alpha+1) i}x^i}{1 + \alpha i} \binom{\alpha i + 1}{i}.
\end{equation}
% For $n=2$, there is only one set $(1,2)$ and therefore
For $m=2$ and the matrix $C=\bigl[\begin{smallmatrix}
\alpha&\beta \\ \beta&\gamma
\end{smallmatrix} \bigr]$
we find $A_{1;C}(l_1, l_2) = \beta l_1$ and then
\begin{equation}
  c^{(1)}_{l_1,l_2} = \frac{(-1)^{(\alpha+1) l_1 + (\gamma+1) l_2} \beta l_1}{(\alpha l_1 + \beta l_2 + 1 )(\beta l_1 + \gamma l_2)}\binom{\alpha l_1 + \beta l_2 + 1}{l_1} \binom{\beta l_1 + \gamma l_2}{l_2} \equiv c^{(1)}_{l_1,l_2}(C).   \label{c1-l1-l2}
\end{equation}
For $m=3$ and the quiver matrix 
\begin{equation}
  C = \begin{bmatrix}
    \alpha & \beta & \delta \\
    \beta & \gamma &\epsilon \\
    \delta & \epsilon & \phi
  \end{bmatrix}
\end{equation}
the polynomial $A_{1;C}(l_1, l_2,l_3)$ reads
\begin{equation}
  A_{1;C}(l_1, l_2, l_3) =  l_1 (\beta \delta l_1 + \beta \epsilon l_2 + \delta \epsilon l_3),
\end{equation}
and then
\begin{align}
\begin{split}
  c^{(1)}_{l_1,l_2,l_3} &= \frac{(-1)^{(\alpha+1) l_1 + (\gamma+1) l_2 + (\phi+1) l_3} (\beta \delta\, l_1^2 + \beta \epsilon\, l_1 l_2 + \delta \epsilon\, l_1 l_3)}{(\alpha l_1 + \beta l_2 + \delta l_3 + 1)(\beta l_1 + \gamma l_2 + \epsilon l_3)(\delta l_1 + \epsilon l_2 + \phi l_3)} \times  \\
  &\quad \times \binom{\alpha l_1 + \beta l_2 + \delta l_3 + 1}{l_1} \binom{\beta l_1 + \gamma l_2 + \epsilon l_3}{l_2}\binom{\delta l_1 + \epsilon l_2 + \phi l_3}{l_3}.
\end{split}
\end{align}

Furthermore, note that the relation (\ref{y-prod-yj}) leads to interesting identities that relate coefficients (\ref{b-l1-lm}) and (\ref{c-1-l1-lm}). For example, for $m=2$, from  
\begin{equation}
  y(x_1, x_2) = y_1(x_1, x_2) y_2(x_1, x_2),
\end{equation}
and the relation (\ref{y2-y1}) we find the following identity for coefficients of $y(x_1, x_2)$ in (\ref{bij}) and $y_1(x_1, x_2)$  in (\ref{c1-l1-l2})
\be
b_{i,j} = \sum_{\substack{k_1 + k_2 = i,\\ l_1 + l_2 = j}} c^{(1)}_{k_1, l_1}(C) c^{(1)}_{l_2, k_2}(C'),
\ee
where $C=\bigl[\begin{smallmatrix}
\alpha&\beta \\ \beta&\gamma
\end{smallmatrix} \bigr]$ and $C'=\bigl[\begin{smallmatrix}
\gamma&\beta \\ \beta&\alpha
\end{smallmatrix} \bigr]$. 
Explicitly, this identity reads
\begin{align}
\begin{split}
  &\frac{\beta i + \beta j + 1} {(\alpha i + \beta j + 1)(\beta i + \gamma j + 1)}\binom{\alpha i + \beta j + 1}{i}\binom{\beta i + \gamma j + 1}{j} =   \\
  & \qquad \qquad = \sum_{\substack{k_1 + k_2 = i,\\ l_1 + l_2 = j}} \frac{\beta k_1}{(\alpha k_1 + \beta l_1 + 1 )(\beta k_1 + \gamma l_1)}\binom{\alpha k_1 + \beta l_1 + 1}{k_1} \binom{\beta k_1 + \gamma l_1}{l_1}  \\ 
  & \qquad \qquad  \qquad \qquad \times \frac{\beta l_2}{(\gamma l_2 + \beta k_2 + 1 )(\beta l_2 + \alpha k_2)}\binom{\gamma l_2 + \beta k_2 + 1}{l_2} \binom{\beta l_2 + \alpha k_2}{k_2}.
  \end{split}
\end{align}
Analogous identities can be easily written down for arbitrary positive integer $m$.

%***************************************************************************************************

\subsection{Quantum curves and A-polynomials for quivers}   \label{ssec-A-quivers}

Quiver generating functions (\ref{P-C}) are built out of quadratic powers of $q$ and $q$-Pochhammers, and depend on variables $x_i$. Therefore they are examples of $q$-holonomic functions, and it is known in general that $q$-holonomic functions satisfy difference equations, which we also refer to as $q$-holonomic equations \cite{doi:10.1080/10236190701264925,Garoufalidis:2016zhf}. It is therefore of interest to determine such difference equations for quiver generating series.

Recall that one important class of $q$-holonomic equations are (generalizations of) quantum A-polynomials for knots, which at the same time are important examples of quantum curves \cite{abmodel,AVqdef}. Furthermore, $q$-difference equations reduce in appropriate limits to differential or algebraic equations. For example quantum A-polynomials for knots reduce to classical A-polynomial algebraic equations, which on one hand encode information about $S^n$-colored knot polynomials for large $n$, and on the other hand capture classical BPS invariants for knots \cite{Garoufalidis:2015ewa}. In case of multiple variables -- which arise for example for knots colored by non-symmetric representations, or for links whose components are independently colored -- higher-dimensional quantum and classical varieties can be considered, such as those discussed in \cite{Aganagic:2013jpa,Gukov:2015gmm}. Note that via the knots-quivers correspondence, quantum A-polynomials for knots at the same time provide difference equations for generating series of quivers associated to knots, in this case with all variables $x_i$ identified with a single variable $x$, as discussed in \cite{Panfil:2018sis}. This provides an interesting example of one class of difference equations for quivers mentioned in the previous paragraph, and motivates us to consider more generally quantum and classical curves and higher-dimensional varieties for quivers, which we also refer to as A-polynomials for quivers. Below we discuss basic properties of such objects, and in the next sections we will take advantage of these results to analyze generating functions for quivers that are associated to branes in strip geometries. 

%Furthermore, recall we have already shown that brane amplitudes for strip geometries take form of generalized $q$-hypergeometric functions (\ref{q-hyper}). It is well known that such functions satisfy $q$-difference equations (generalized $q$-hypergeometric equations), which therefore can be thought of as quantum curves for this class of topological string amplitudes. On the other hand, in section \ref{sec-top-quivers} we show that brane amplitudes in question can be identified with motivic generating series for appropriately chosen quivers. This motivates us to study in general quantum curves and varieties, and their classical analogs, for quivers. 

Let us introduce operators $\hat{x}_i$ and $\hat{y}_i$ that satisfy the relation
\begin{equation}
  \hat{x}_i\hat{y}_j = q^{\delta_{ij}}\hat{y}_j \hat{x}_i,
\end{equation}
and consider a $q$-series $\psi(x_1,\ldots, x_m)$ that depends on variables $x_i$, on which the above operators act as
\begin{align}
\begin{split}
\hat{x}_i \psi(x_1, \dots, x_n) & = x_i \psi(x_1, \dots, x_n),\\
\hat{y}_i \psi(x_1, \dots, x_n) &= \psi( x_1, \dots, x_{i-1}, q x_i, x_{i+1}, \dots, x_n).
\end{split}
\end{align}
In general we may ask whether  the following set of finite difference equations is satisfied
\begin{equation}
\hat{A}_i(\hat{x}_1, \dots, \hat{x}_m, \hat{y}_1,\ldots,\hat{y}_m) \psi(x_1, \dots, x_m) = 0,\qquad i=1,\ldots,m.  \label{Ai-hat-quiver}
\end{equation}
Such equations would define a higher-dimensional quantum variety, which in the classical limit $q\to 1$ would reduce to a classical variety defined by a set of algebraic equations $A_i(x_1,\ldots,x_m,y_1,\ldots,y_m)=0$ \cite{Aganagic:2013jpa,Gukov:2015gmm}. 

Consider now a $q$-series $\psi(x)=P_C(x_1,\ldots,x_m)$ that takes form of the quiver generating functions (\ref{P-C}). It turns out that in this case we can identify separate equations that involve only a single $\hat{y}_i$ operator 
\begin{equation}
\hat{A}_i(\hat{x}_1, \dots, \hat{x}_m, \hat{y}_i) P_C(x_1, \dots, x_m) = 0,\qquad i=1,\ldots,m.   \label{A-hat-PC}
\end{equation}
In this case in the classical limit we get a set of equations 
\be
A_i(x_1,\ldots,x_m,y_i)=0.    \label{A-hat-PC-class}
\ee 
These equations can be solved for $y_i=y_i(x_1,\ldots,x_m)$, which are the same functions that arise in the partial classical limit (\ref{y-j}). The functions $y_i(x_1,\ldots,x_m)$ can be also determined from the analysis of the asymptotic expansion of the motivic generating series (\ref{P-C}). Indeed, taking advantage of the expansion of the $q$-Pochhammer symbol
\begin{equation}
  (x;q)_{d} \simeq e^{ \frac{1}{\hbar}({\rm Li}_2 (x) - {\rm Li_2}(q^d x)) + \ldots},   \label{qPoch-asymptotics}
\end{equation}
and approximating the sums over $d_i$ in (\ref{P-C}) by integrals over $z_i = e^{\hbar d_i}$, we get
\begin{equation}
  P_C(x_1, \dots, x_n) \simeq \int%_{1}^{\infty} 
\frac{dz_1\cdots dz_m}{ z_1 \cdots z_m} \exp\Big(\frac{1}{\hbar} W(x,z)\Big),   \label{P-C-int}
\end{equation}
with the potential
\begin{equation}
  W(x,z) = \frac{1}{2}\sum_{i,j=1}^m C_{i,j}\log z_i  \log z_j + \sum_{i=1}^m\Big( \log z_i \log x_i + {\rm Li}_2(z_i) - {\rm Li}_2(1) + i \pi  C_{i,i} \log z_i \Big).
\end{equation}
In $\hbar \rightarrow 0$ limit we can evaluate integrals in (\ref{P-C-int}) using the saddle point method, by finding stationary points of the potential $\partial_{z_i} W(x,z)= 0$. After exponentiating, these saddle point equations take form
\begin{equation}
  1 - z_i = (-1)^{C_{i,}} \, x_i \prod_{j=1}^n z_j^{C_{i,j}}, \qquad i=1, \dots,m,
\end{equation}
and we denote their solutions by $\bar{z}=(\bar{z}_i)_{i=1,\ldots,m}$. It follows that the partial classical limit (\ref{y-j}) can be also evaluated as
\begin{equation}
  y_i(x_1, \dots, x_n)  = e^{\partial_{x_i} W(x,\bar{z})} = \bar{z}_i.
\end{equation}
Moreover, the complete classical limit (\ref{y-x1-xm}) is simply
\begin{equation}
  y(x_1, \dots, x_n) =e^{\sum_{i=1}^m \partial_{x_i} W(x,\bar{z})}  = \prod_{j=1}^m \bar{z}_i.
\end{equation}
The last two equations imply that $y(x_1, \dots, x_m)$ factorizes into
\begin{equation}
  y(x_1, \dots, x_n) = \prod_{i=1}^m y_i(x_1, \dots, x_m),
\end{equation}
in agreement with (\ref{y-prod-yj}).

In what follows we will analyze quantum and classical A-polynomials for those quivers, which we will associate to strip geometries. We will discuss the relation of these A-polynomials to quantum and classical mirror curves for strip geometries. Moreover, in view of the relation of partition functions for branes in strip geometries to generalized $q$-hypergeometric functions (\ref{q-hyper}), we will also see that A-polynomials for corresponding quivers are related to $q$-hypergeometric equations (\ref{A-hat-psi-f}) and their limits.

%***************************************************************************************************
%***************************************************************************************************

\section{Topological strings and quivers}   \label{sec-top-quivers}    \label{sec-top-quivers}

In this section we derive the main result of this work, which is the statement that to a brane in a strip geometry one can associate the corresponding quiver, such that various characteristics of this brane (its partition function, BPS invariants, etc.) are encoded in the moduli space of representations of the corresponding quiver. We also propose the interpretation of vertices of this quiver, as corresponding to discs that represent open BPS states associated to a given strip geometry. Furthermore, we relate quantum and classical mirror curves to A-polynomials for quivers, derive explicit expressions for classical BPS invariants for an arbitrary strip geometry, and discuss constraints on the structure of BPS invariants for strip geometries that follow from the quiver interpretation.

%***************************************************************************************************

\subsection{Brane amplitudes as quiver generating functions}

To start with, recall that we derived the following expression for the brane generating function in a strip geometry (\ref{psi-hyper})
\be
\psi_f(x) = \sum_{n=0}^{\infty} \big((-1)^{n} q^{n(n-1)/2}  \big)^{f+1}  \frac{ x ^n}{(q;q)_n} \frac{(\alpha_1;q)_n (\alpha_2;q)_n\cdots (\alpha_r;q)_n}{(\beta_1;q)_n (\beta_2;q)_n\cdots (\beta_s;q)_n},    \label{psi-f-bis}
\ee
where $x$ is the open string generating parameter, and $\alpha_i$ and $\beta_j$ are appropriate products of K{\"a}hler parameters $Q_k$ that characterize the underlying strip geometry. This amplitude is nothing but a simple generalization of the definition of the $q$-hypergeometric function, which arises from the above formula once the framing $f=s-r$ is chosen (\ref{q-hyper}). We now show that this generating function can be rewritten in the form of the motivic quiver generating function (\ref{P-C}). To this end note that the following expansions of the quantum dilogarithm and its inverse
\be
(\alpha;q)_{\infty} = \sum_{i=0}^{\infty} \frac{(-1)^i q^{i(i-1)/2} \alpha^i}{(q;q)_i},\qquad \quad
\frac{1}{(\alpha;q)_{\infty}} = \sum_{i=0}^{\infty} \frac{\alpha^i}{(q;q)_i},    \label{qPoch-infty-series}
\ee
enable to rewrite $q$-Pochhammers $(\alpha_i;q)_n$ and their inverses $(\beta_j;q)_n^{-1}$ in (\ref{psi-f-bis}) in the form
\begin{align}
\begin{split}
  (\alpha;q)_n &= \frac{(\alpha;q)_{\infty}}{(\alpha q^n;q)_{\infty}} = \sum_{i,j} (-q^{-1/2}\alpha)^i\alpha^j \frac{q^{i^2/2+jn}}{(q;q)_i (q;q)_j}, \\ 
  \frac{1}{(\beta;q)_n} &= \frac{(\beta q^n;q)_{\infty}}{(\beta;q)_{\infty}} = \sum_{i,j} (-q^{-1/2}\beta)^i\beta^j \frac{ q^{i^2/2+in}}{(q;q)_i (q;q)_j}.  \label{expansions}
\end{split}
\end{align}
Expanding all $q$-Pochhammers in (\ref{psi-f-bis}) in this way and comparing the resulting expression with (\ref{P-C}), we find that the brane generating function can be written in the form of the quiver generating series
\begin{equation}
\psi_f(x) = P_C(q^{-(f+1)/2}x, q^{-1/2}\alpha_1, \alpha_1, \dots, q^{-1/2}\alpha_r, \alpha_r, q^{-1/2}\beta_1, \beta_1, \dots, q^{-1/2}\beta_s, \beta_s), \label{psi-f-quiver}
\end{equation}
for a quiver whose structure is encoded by the symmetric matrix of size $2(r+s)+1$ 
%\begin{equation}
%  C = \begin{bmatrix}
%    f+1 & 0 & 1 & \cdots & 0 & 1 & 1 & 0 & \cdots & 1 & 0 \\
%    0& 1 & 0 & \cdots & & & & & & & 0 \\
%    1& 0 & 0 & 0 &\cdots & & & & & & 0 \\
%    \vdots& & \vdots & 1 & 0 & \cdots & & & & & 0 \\
%    & & & \vdots & 0 & 0 & \cdots & & & & 0 \\ 
%    & & & & & \ddots& & & & & \vdots       \\
%     & & & & & & & & &  &  \\
%    \vdots  & & & & & & & & \ddots & & \vdots \\
%     1 & & & & & & & & & 1 & 0 \\
%    0& & & & & & & & & & 0
%  \end{bmatrix}   \label{C-psi-f}
%\end{equation}
\begin{equation}
  C = \left[\begin{array}{c|ccccc|ccccc}
              f+1 & 0 & 1 & \dots & 0 & 1 & 1 & 0 & \dots & 1 & 0 \\ \hline
              0 & 1 & 0 & \dots & 0 & 0 & 0 & & \dots  & & 0 \\
              1 & 0 & 0 & \dots & 0 & 0 & 0 & & \dots & & 0 \\
              \vdots & & & \ddots & & & & & \ddots & &  \\
              0 & 0 & 0 & \dots & 1 & 0 & 0 & & \dots  & & 0 \\
              1 & 0 & 0 & \dots & 0 & 0 & 0 & & \dots  & & 0 \\ \hline
              1 & 0 & & \dots & & 0 & 1 & 0 & \dots & 0 & 0 \\
              0 & 0 & & \dots & & 0 & 0 & 0 & \dots & 0 & 0 \\
              \vdots & & & \ddots & & & & & \ddots & &  \\
              1 & 0 & & \dots & & & 0 & 0 & \dots & 1 & 0 \\
              0 & 0 & & \dots & & 0 & 0 & 0 & \dots & 0 & 0
        \end{array}\right].    \label{C-psi-f}
\end{equation}
In more detail, this matrix has non-zero entries only in the first row, the first column, and along the diagonal. The first row, and analogously the first column, consist of the first entry $f+1$, followed by $r$ pairs of entries $(0,1)$, and then $s$ pairs of entries $(1,0)$. The diagonal consist of the first entry $f+1$, followed by $r+s$ pairs of entries $(1,0)$. The structure of this matrix simply follows from the quadratic powers of $q$ in (\ref{expansions}) and the framing factor in (\ref{psi-f-bis}).

In particular, for $f=s-r$ we find the following quiver representation of the generalized $q$-hypergeometric function
\begin{align}
\begin{split}
&\pphiq{r}{s}{\alpha_1,\alpha_2, \dots, \alpha_r}{\beta_1, \dots, \beta_r}{q}{x} = \\
& \qquad \quad = P_C(q^{(r-s-1)/2}x, q^{-1/2}\alpha_1, \alpha_1, \dots, q^{-1/2}\alpha_r, \alpha_r, q^{-1/2}\beta_1, \beta_1, \dots, q^{-1/2}\beta_s, \beta_s). \label{qHyper_quiver}
\end{split}
\end{align}
This form implies new interesting properties of generalized $q$-hypergeometric functions, and so also ordinary generalized hypergeometric functions. 

Furthermore, note that the size of the above matrix $2(r+s)+1$, which is equal to the number of vertices in the quiver, indicates the interpretation of these vertices as corresponding to discs associated with each strip geometry, which represent open BPS states. Recall that to each local $\mathbb{P}^1$ one can associate two local discs wrapping its two hemispheres -- for the resolved conifold they are captured by two non-zero BPS invariants, and they also represent two HOMFLY-PT homology generators of the unknot in knot theory interpretation. Analogously, a single brane in $\mathbb{C}^3$ captures just one disc, representing a single BPS state. A strip geometry labeled by a pair $(r,s)$ consists of $r+s$ local $\mathbb{P}^1$'s, so together with one additional disc associated to the brane it then indeed encodes $2(r+s)+1$ fundamental discs representing open BPS states, in agreement with the size of the matrix $C$. Moreover, the fact that changing framing changes the number of loops only at one vertex, which corresponds to the entry $C_{1,1}$ of the quiver matrix and can be associated to the brane under consideration (and not any other $\mathbb{P}^1$), supports this interpretation. Note that a similar identification of vertices of a quiver corresponding to a knot is proposed in the context of knots-quivers correspondence in \cite{EKL}.

We note that we can also represent brane amplitudes (\ref{psi-f-bis}) in terms of quivers of smaller size. When rewriting factors $(\alpha;q)_n$ and $(\beta;q)_n^{-1}$ in (\ref{expansions}) we can keep the factors $(\alpha;q)_{\infty}$ and $(\beta,q)_{\infty}$, which equivalently arise from partial resummations in $\psi_f(x)$. It follows that
\be 
  \psi_f(x) = 
  \frac{\prod_{j=1}^r (\alpha_j;q)_{\infty}}{\prod_{j=1}^s (\beta_j;q)_{\infty}} \times P_{C'}(q^{(r-s-1)/2} x, \alpha_1, \dots, \alpha_r, q^{-1/2}\beta_1, \dots, q^{-1/2}\beta_s), \label{qHyper_quiver2}
\ee
where a quiver matrix $C'$ is of size $(r+s+1)$, and it is obtained from $C$ by removing all rows and columns (other than the first one) whose first entry is zero:
%\begin{equation}
%  C' = \begin{bmatrix}
%    f+1 & 1 & \cdots & 1 & 1 & \cdots & 1  \\
%    1& 0 & 0 & \cdots  & & & 0 \\
%    \vdots& & \ddots &   &   & & \vdots \\
%    & & & & &  &   \\
%     \vdots & & & & & \ddots & \vdots  \\
%    1 & & & & & &  1  \\
%  \end{bmatrix}   \label{C-prim-psi-f}
%\end{equation}
\begin{equation}
  C' = \left[\begin{array}{c|ccc|ccc}
              f+1 & 1 & \dots &  1 & 1 & \dots & 1  \\ \hline
              1 & 0 & \dots & 0 & 0 & \dots & 0 \\
              \vdots & & \ddots & & & \ddots &  \\
              1 & 0 & \dots & 0 & 0 & \dots & 0 \\ \hline
              1 & 0 & \dots & 0 & 1 & \dots & 0 \\
              \vdots & & \ddots & & & \ddots &  \\
              1 & 0 & \dots & 0 & 0 & \dots & 1 \\
        \end{array}\right].      \label{C-prim-psi-f}
\end{equation}

In the rest of this section we discuss several consequences of the relation between topological string amplitudes for strip geometries and quivers. However, before proceeding, let us also stress, that while the above relation is analogous to the knots-quivers correspondence \cite{Kucharski:2017poe,Kucharski:2017ogk}, there are also several important differences. First, the brane partition function $\psi_f(x)$ depends on the modulus $x$, which is identified only with one generating parameter $x_1$ of the quiver generating function (\ref{psi-f-quiver}), while in the knots-quivers correspondence all quiver generating parameters $x_1,\ldots,x_m$ are proportional to $x$. On the other hand, in the present context quiver generating parameters $x_2,\ldots,x_m$ are identified with combinations of a number of closed string moduli encoded in $\alpha_i$ and $\beta_j$, while in the knots-quivers correspondence only one additional variable $a$ of HOMFLY-PT polynomials had to be taken into account. As already mentioned, for strip geometries the change of framing changes the number of loops only at one vertex, while for knots it changes by the same amount the number of loops at each vertex of the corresponding quiver. Because of these differences, certain aspects of the relation between strip geometries and quivers are different than in the knots-quivers correspondence.

%***************************************************************************************************

\subsection{Quantum curves and A-polynomials}

As we just stressed, and as seen in (\ref{psi-f-quiver}), for strip geometries only the first variable $x_1$ in the motivic generating series is identified with the brane modulus $x$. Therefore $\psi_f(x)$ must be annihilated by the partial $\widehat{A}_1$ operator in (\ref{A-hat-PC}), with appropriate identification of other parameters. On the other hand, we have already shown that brane partition functions for strip geometries (\ref{psi-f-bis}) are annihilated by the operators $\widehat{A}(\hat x, \hat y)$ of the form (\ref{A-hat-psi-f}). This means that these two operators, with the identification of parameters as in (\ref{psi-f-quiver}), must be equal
\be
\widehat{A}(\hat x, \hat y) = \widehat{A}_1 (q^{-(f+1)/2}\hat x, q^{-1/2}\alpha_1, \alpha_1, \dots, q^{-1/2}\alpha_r, \alpha_r, q^{-1/2}\beta_1, \beta_1, \dots, q^{-1/2}\beta_s, \beta_s,\hat y),  \label{A-A1-hats}
\ee
and in consequence equations defining mirror curves (\ref{A-hat-psi-f-class}) also take form
\be
A(x, y) = A_1 (x, \alpha_1, \alpha_1, \dots, \alpha_r, \alpha_r, \beta_1, \beta_1, \dots, \beta_s, \beta_s,y),     \label{A-A1-class}
\ee
with $A_1(x_1,\ldots,x_m,y_1)$ given in (\ref{A-hat-PC-class}). In section \ref{sec-examples} we will illustrate in various examples that this is indeed the case.

Moreover, in view of our results concerning the partial classical limit introduced in section \ref{ssec-partial}, we can now write down explicit and exact expressions for the coefficients of the series 
\be
y=y(x) = \sum_{i=0}^{\infty} c_i x^i
\ee
that is a solution of the mirror curve equation $A(x,y)=0$ in (\ref{A-hat-psi-f-class}), for an arbitrary strip geometry. Indeed, the coefficients of the function that solves the partial equation $A_1(x_1,\ldots,x_m,y_1)=0$ in (\ref{A-hat-PC-class}) are given in (\ref{c-1-l1-lm}). We can now determine these coefficients for an arbitrary matrix $C$ in (\ref{C-psi-f}), or equivalently $C'$ in (\ref{C-prim-psi-f}), corresponding to a given strip geometry. In view of the identification (\ref{A-A1-class}), and -- using the form $C'$ in (\ref{C-prim-psi-f}) -- identifying quiver variables as 
\be
x_1=x,\qquad x_2=\alpha_1,\ldots,x_{1+r}=\alpha_r, \qquad x_{2+r}=\beta_1,\ldots,x_{1+r+s}=\beta_s,    \label{x-alpha-beta}
\ee
and denoting $|l| = \sum_{j} l_j$, we find
\begin{align}
\begin{split}
  c_{i} & =  \sum_{l_1, \dots, l_r}\sum_{k_1, \dots, k_s} \frac{(-1)^{f i}  }{1 + (f+1)i + |l| + |k|}\binom{1 + (f+1)i + |l| + |k|}{i} \times  \\
  &\qquad\qquad\quad \times \prod_{j=1}^r (-1)^{l_j}  \binom{i}{l_j} \alpha_j^{l_j}\, \prod_{j=1}^s \frac{i}{i+k_j}  \binom{i+k_j}{k_j} \beta_j^{k_j}.    \label{y-c_i}
  \end{split}
\end{align}

%***************************************************************************************************

\subsection{BPS invariants and their structure}

The fact that brane partition functions can be expressed in terms of motivic generating functions for quivers has important consequences. First, the product decomposition of the brane partition function (\ref{Z-open-x}) into quantum dilogarithms is analogous to the product decomposition of the quiver generating function (\ref{PQx-Omega}). It follows that open BPS (Ooguri-Vafa) invariants $N_{n,\beta,j}$ can be expressed as combinations, with integer coefficients, of motivic Donaldson-Thomas invariants $\Omega_{d_1,\ldots,d_m;j}$. This immediately proves that open BPS invariants for strip geometries are integer -- and this is an important conclusion in itself.

Moreover, using the results from section \ref{ssec-partial} we can write down explicit expressions for classical BPS invariants for an arbitrary strip geometry. For a quiver $C'$ in (\ref{C-prim-psi-f}), with the same identification of parameters as in (\ref{x-alpha-beta}), the product decomposition (\ref{y_i-Omega}) takes form
\begin{equation}
  y(x, \alpha_1, \dots, \alpha_r, \beta_1, \dots, \beta_s) = \prod_{(n, \mathbf{l}, \mathbf{k})>0} \left(1 - x^n \alpha_1^{l_1} \cdots \alpha_r^{l_r} \beta_1^{k_1}\cdots \beta_s^{k_s} \right)^{n\Omega_{n,\mathbf{l}, \mathbf{k}}},   \label{y-x-alpha-beta}
\end{equation}
where we now denote the sets of indices as $\mathbf{l}=(l_1,\ldots,l_r), \mathbf{k}=(k_1,\ldots,k_s)$. The classical Donaldson-Thomas invariants (\ref{Omega-numerical}) 
\begin{equation}
  \Omega_{n,\mathbf{l}, \mathbf{k}} = \sum_j (-1)^j \Omega_{n,\mathbf{l}, \mathbf{k};j},
\end{equation}
can be expressed through the coefficients (\ref{y-c_i}). To this end we compute the logarithm of (\ref{y-x-alpha-beta}), on one hand, as
\begin{equation}
  \log y(x) =  \sum_{n, \mathbf{l}, \mathbf{k}} \Big(\sum_{i|{\rm gcd}(n, \mathbf{l}, \mathbf{k})}\frac{n}{i^2} \Omega_{n/i, \mathbf{l}/i, \mathbf{k}/i}\Big) x^n \alpha_1^{l_1} \cdots \alpha_r^{l_r}\beta_1^{k_1}\cdots \beta_s^{k_s}.     \label{log-y-alpha-beta}
\end{equation}
On the other hand, the same logarithm arises as specialization of (\ref{log-y_1}) to (\ref{C-prim-psi-f}) and (\ref{A-A1-class})
\begin{equation}
  \log y(x) = \sum_{n, \mathbf{l}, \mathbf{k}} d_{n, \mathbf{l}, \mathbf{k}} \,x^n \prod_{j=1}^r \alpha_j^{l_j} \prod_{j=1}^s \beta_j^{k_j},    \label{log-y-alpha-beta-bis}
\end{equation}
so that the coefficients (\ref{log-y_1-d}) take form
\begin{equation}
  d_{n, \mathbf{l}, \mathbf{k}} = \frac{(-1)^{f n}  }{(f+1)n + |l| + |k|}\binom{(f+1)n + |l| + |k|}{n} \prod_{j=1}^r (-1)^{l_j}\binom{n}{l_j} \prod_{j=1}^s \frac{n}{n+k_j}\binom{n+k_j}{k_j}.
\end{equation}
Comparing coefficients in (\ref{log-y-alpha-beta}) and (\ref{log-y-alpha-beta-bis}), or equivalently specializing (\ref{Omega-d1-dm}) to (\ref{C-prim-psi-f}) and (\ref{A-A1-class}), we find
\begin{align}
\begin{split}
  \Omega_{n, \mathbf{l}, \mathbf{k}} &=  \frac{1}{n} \sum_{i|{\rm gcd}(n, \mathbf{l}, \mathbf{k})} \frac{\mu(i)}{i} d_{n/i, \mathbf{l}/i,\mathbf{k}/i}= \\
  &= - \frac{1}{n} \sum_{i|{\rm gcd}(n, \mathbf{l}, \mathbf{k})}\frac{(-1)^{f n/i} \mu(i)  }{(f+1)n + |l| + |k|}\binom{\left((f+1)n + |l| + |k|\right)/i}{n/i} \times \\
  &\quad \times\prod_{j=1}^r (-1)^{l_j/i}\binom{n/i}{l_j/i} \prod_{j=1}^s \frac{n}{n+k_j}\binom{(n+k_j)/i}{k_j/i}.    \label{Omega-nlk-classical}
\end{split}
\end{align}
This is an explicit expression for open BPS invariants of an arbitrary strip geometry, in arbitrary framing. Note that this formula gives a large set of integrality statements -- as $\Omega_{n, \mathbf{l}, \mathbf{k}} $ are classical Donaldson-Thomas invariants for the quiver (\ref{C-prim-psi-f}) we know that they are integer, despite the factor of $1/n$ and other denominators. This vastly generalizes analogous results for the framed unknot, or equivalently a brane in $\mathbb{C}^3$ or resolved conifold, presented in \cite{Garoufalidis:2015ewa,Luo:2016oza}. It would also be interesting to provide a purely number theoretic proof of integrality of (\ref{Omega-nlk-classical}), generalizing the proof for the extremal unknot invariants (or equivalently a brane in $\mathbb{C}^3$) in \cite{Basor:2017qxy}; and it is of interest to relate these integrality statements to the formalism of \cite{Schwarz:2013zua}.

Note that we determined classical BPS invariants (\ref{Omega-nlk-classical}) upon the analysis of the function (\ref{y-x-alpha-beta}), which satisfies mirror curve equation (\ref{A-A1-class}). As for strip geometries we also know the form of the quantum curve (\ref{A-hat-psi-f}), in principle one could construct statistical models for quantum BPS states and identify corresponding invariants, following the formalism presented in \cite{Kucharski:2016rlb}. 

Furthermore, we can get some insight into the structure of quantum BPS states more directly. The fact that brane partition functions for strip geometries take form similar (just ``framed'') to generalized $q$-hypergeometric functions (\ref{psi-f-bis}), for which the limit (\ref{tilde-psi}) exists, already imposes non-trivial constraints on the form of BPS invariants and motivic Donaldson-Thomas invariants of the corresponding quiver. Indeed, brane partition functions in the decomposition (\ref{Z-open-x}) or (\ref{PQx-Omega}) are products of quantum dilogarithms. In view of the asymptotics (\ref{qPoch-asymptotics}), in the limit $q=e^{\hbar}\to 1$ these functions behave as 
\be
\psi_f(x)\sim \exp\Big(\big(\frac{1}{\hbar}\sum \textrm{Li}_2(x^{\#} \prod_i \alpha_i^{\#} \prod_j \beta_j^{\#}\big) + \mathcal{O}(1) + \ldots \Big),     \label{psi-exp}
\ee
where $\#$ denote certain powers. At first sight this is a singular behavior. Nonetheless, we know that the non-singular limit (\ref{tilde-psi}) exists, in which $\widetilde{\psi}=1+\mathcal{O}(x)$. This means, that the singular $\frac{1}{\hbar}$ behavior in (\ref{psi-exp}) must cancel. Such a cancellation may arise in two ways. First, this may follow from the rescaling (\ref{rescale}), if only $1+s-r\neq 0$; in this case $(q-1)=\hbar+\ldots$, and altogether after the rescaling $x$ may be multiplied by a non-zero power of $\hbar$. We can then expand $\textrm{Li}_2(\hbar^{c_1}x^{c_2})=\hbar^{c_1}x^{c_2}+\ldots$, and if here $c_1=1$, we get a cancellation with the overall $\frac{1}{\hbar}$ in (\ref{psi-exp}), and we get a non-trivial contribution; on the other hand, for $c_1>1$ we will get no contribution in $\hbar\to 0$ limit.

The second possibility to cancel $\frac{1}{\hbar}$ behavior in (\ref{psi-exp}) arises when an intricate relation between $\Omega_{d_1,\ldots,d_m;j}$ holds, such that several dilogarithm terms cancel each other in the limit $\hbar\to 1$. In particular such a behavior must happen when $1+s-r=0$ in (\ref{rescale}), as in this case $x$ cannot get accompanied by any factor of $\hbar$, and the only possibility to cancel $\frac{1}{\hbar}$ behavior is to cancel dilogarithm terms among themselves. 

Moreover, additional constraints on BPS invariants can be deduced from the form of the differential equation (\ref{A-tilde-psi-f}) that arises in the above limit.

As an illustration of the above statements, consider a quiver with one vertex and $\alpha$ loops, determined by the matrix $C=[\alpha]$, and suppose that its motivic generating function has a product decomposition (\ref{PQx-Omega}) of the form
\begin{equation}
  P_{\alpha}(x) = \sum_{d=0}^{\infty} \frac{(-q^{1/2})^{\alpha d^2}}{(q;q)_d} x^d = \prod_{d>0}\prod_{j \in \mathbb{Z}}(x^dq^{(j+1)/2};q)_{\infty}^{(-1)^{j+1} \Omega_{d;j}}.  \label{Palpha}
\end{equation}
From the asymptotics 
\be
(x;q)_{\infty} =\exp\Big(\frac{1}{\hbar} \textrm{Li}_2(x) + \frac{1}{2}\log(1-x) + \mathcal{O}(\hbar) \Big)
\ee
(which also implies (\ref{qPoch-asymptotics})) it follows that
\begin{equation}
  P_{\alpha}(x) = \exp\Big(\sum_{d>0}\sum_{j \in \mathbb{Z}} (-1)^{j+1}\Omega_{d;j} \big(\frac{1}{\hbar}{\rm Li}_2(x^d) - \frac{j}{2}\log(1-x^d) + \mathcal{O}(\hbar) \big)\Big). \label{Palpha-exp}
\end{equation}
For this limit to exist, there are two possibilities. First, if $x$ would be rescaled as $x\to (q-1)x=\hbar x+\ldots$, in the limit the only contribution would arise from the dilogarithm terms for $d=1$, and we would get the exponential function
\begin{equation}
\widetilde{P}_{\alpha}(x) = \lim_{\hbar\to 0}  P_{\alpha}((q-1)x) = \exp\Big(x \sum_{j\in\mathbb{Z}}(-1)^{j+1} \Omega_{1;j}\Big).  \label{tildePalpha}
\end{equation}
Furthermore, the generating series (\ref{Palpha}) satisfies the difference equation easily obtained from (\ref{A-hat-psi-f})
\be
\big(1 -\hat{y} - (-1)^{\alpha} q^{\alpha/2} \hat{x} \hat{y}^{\alpha} \big) P_{\alpha}(x) = 0.  \label{Ahat-Palpha}
\ee
Writing $\hat y=e^{\hbar x\partial_x}$, rescaling $x\to (q-1)x=\hbar x+\ldots$, and taking $\hbar\to 0$ limit, this equation reduces to 
\be
\big(\partial_x + (-1)^{\alpha}\big) \widetilde{P}_{\alpha}(x) = 0.
\ee
The solution of this last equation is $e^{-(-1)^{\alpha} x}$. It then follows that the coefficient in the exponent in (\ref{tildePalpha}) must satisfy 
\be
\sum_{j\in\mathbb{Z}}(-1)^{j+1} \Omega_{1;j}=-(-1)^{\alpha}.    \label{Omega-constraint}
\ee
This imposes an additional non-trivial condition on coefficients $\Omega_{1;j}$.

On the other hand, if we assume that the limit $\hbar\to 0$ exists but $x$ is not rescaled, we must require that all dilogarithm terms in (\ref{Palpha-exp}) cancel among each other
\begin{equation}
  \sum_{j\in\mathbb{Z}} (-1)^{j+1} \Omega_{d;j} = 0 \qquad \forall\, d > 0. \label{no_rescalling_1node}
\end{equation}
In this case in the $\hbar\to 0$ limit we would get the result of the form
\begin{equation}
\lim_{\hbar\to 0}  P_C(x) = \prod_{d>0} (1-x^d)^{-\frac{1}{2}\sum_{j\in\mathbb{Z}}(-1)^{j+1} j\Omega_{d;j}}.
\end{equation}
However in the limit $\hbar\to 0$, without rescaling of $x$, the equation (\ref{Ahat-Palpha}) does not reduce to a meaningful differential equation. Therefore, if we insist the such differential equation should exist, this indicates that $x$ must be rescaled, as analyzed first.

As another example we consider a quiver with 3 vertices, encoded in a matrix $C$. Suppose we identify the generating parameters as $x_1=x$, $x_2=q^{a_1}$, and $x_3=q^{a_2}$. We then find
\begin{align}
\begin{split}
  P_C(x, q^{a_1}, q^{a_2}) &= \exp\Big(\sum_{d_1,d_2,d_3}\sum_{j \in \mathbb{Z}} (-1)^{j+1}\Omega_{d_1,d_2,d_3;j} \times \\
  &\qquad  \times
  \big(\frac{1}{\hbar}{\rm Li}_2(x^{d_1}) - (j + a_1 d_2 + a_2 d_3)\log(1-x^{d_1}) + \mathcal{O}(\hbar) \big)\Big).
\end{split}
\end{align}
If we do not rescale $x$, in order to avoid a singular behavior we must impose the condition
\begin{equation}
\sum_{d_2,d_3}\sum_{j \in \mathbb{Z}}  (-1)^{j}\Omega_{d_1,d_2,d_3;j} = 0 \qquad \forall  d_1,    \label{Omega-constraint-3}
\end{equation}
and then in the limit $q\to 1$ we get
\begin{equation}
 \lim_{q\to 1} P_C(x, q^{a_1}, q^{a_2}) = \prod_{d_1>0} (1 - x^{d_1})^{\sum_{d_2,d_3}\sum_{j \in \mathbb{Z}} (-1)^{j}\Omega_{d_1,d_2,d_3;j} (j + a_1 d_1 + a_2 d_2)}.  \label{PC-a1a2}
\end{equation}

%***************************************************************************************************
%***************************************************************************************************

\section{Examples}   \label{sec-examples}

In this section we illustrate various results found above in several examples of strip geometries. It is convenient to label these examples by a pair of integers $(r,s)$, which indicates their relation to generalized $q$-hypergeometric functions $_r\phi_s[\, \cdots \, ;q,x]$. We identify corresponding quivers, BPS invariants, quantum varieties and A-polynomials, and analyze their classical limits.

%***************************************************************************************************

\subsection{$\mathbb{C}^3$ geometry ($r=0, s=0$)}

\begin{figure}[h]
\begin{center}
\includegraphics[width=0.25\textwidth]{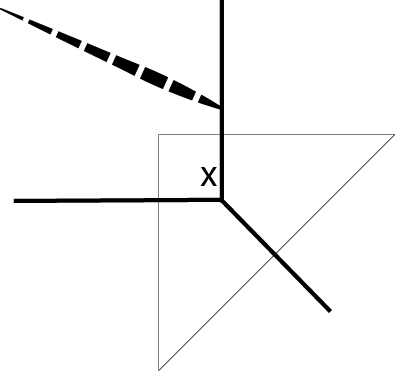} 
\caption{$\mathbb{C}^3$.}  \label{fig-strip-C3}
\end{center}
\end{figure}

To start with we consider $\mathbb{C}^3$, the simplest toric geometry, whose diagram is shown in fig. \ref{fig-strip-C3}. It is well known that brane amplitudes in this case encode extremal colored HOMFLY-PT invariants of the unknot, and the corresponding quiver consists of one vertex and an arbitrary number of loops, which corresponds to the choice of framing $f$ \cite{Rei12,Kucharski:2017poe,Kucharski:2017ogk}. The quiver matrix (\ref{C-psi-f}) is simply $C=[f+1]$ for arbitrary framing $f$, and the motivic quiver generating function (which is just (\ref{Palpha}) with $\alpha=f+1$), and the brane partition function (\ref{psi-f-quiver}), respectively take form
\begin{equation}
  P_C(x) = \sum_d (-1)^{(f+1)d} x^d \frac{q^{(f+1)d^2/2}}{(q;q)_d}, \qquad \quad \psi_f(x)=P_C(q^{-(f+1)/2} x).
\end{equation}
Even though this is the simplest quiver generating function, for generic values of $f$ it encodes an infinite number of motivic Donaldson-Thomas invariants (\ref{PQx-Omega}). However for a special choice of $f=0$, taking advantage of (\ref{qPoch-infty-series}), the above sum is an expansion of a single quantum dilogarithm, and comparing with (\ref{PQx-Omega})
\be
P_{C=[1]}(x) = (q^{1/2}x;q)_{\infty}  \equiv \prod_{d,j} \prod_{k=1}^{\infty} (1-x^d q^{k+(j-1)/2})^{(-1)^{j+1}\Omega_{d;j}}    \label{P-C-1-x}
\ee
it follows that it encodes a single motivic Donaldson-Thomas invariant  $\Omega_{1;0}= -1$; note that its value is consistent with the constraint (\ref{Omega-constraint}). At the same time, $\psi_{f=0}(x)$ is the simplest example of the $q$-hypergeometric function (\ref{q-hyper})
\be
\psi_{f=0}(x) = (x;q)_{\infty} %= \pphiq{0}{0}{\{\}}{\{\}}{q}{x} 
= \, _0\phi_0[\, \cdot\, ;q,x].
\ee

For $\mathbb{C}^3$ geometry the quantum curve operator that annihilates the brane amplitdue (\ref{A-hat-psi-f}) takes form
\begin{equation}
  \hat{A}(\hat{x}, \hat{y}) = 1-\hat{y} + (-1)^{f} \hat{x}\hat{y}^{f+1} ,\qquad \quad \hat{A}(\hat{x}, \hat{y}) \psi_f(x) = 0,
\end{equation}
and it reduces to the $q$-hypergeometric equation for $f=0$
\begin{equation}
 (1-x)\psi_{f=0}(qx) - \psi_{f=0}(x) = 0.
\end{equation}

In the classical limit the quantum curve reduces to the classical mirror curve
\be
A(x,y) = 1 - y + (-1)^f  x y^{f+1} = 0,
\ee
and the solution of this equation for $y$ immediately follows from (\ref{y-c_i})
\be
y = y(x) = \sum_{i=0}^{\infty} \frac{(-1)^{fi}}{1+(f+1)i}{1+(f+1)i \choose i} x^i,
\ee
which nicely illustrates the power of the partial classical limit that led to (\ref{y-c_i}). Furthermore, in this case classical BPS invariants (\ref{Omega-nlk-classical}) take form
\be
\Omega_n = -\frac{1}{(f+1) n^2}  \sum_{i|n}  \mu(i) (-1)^{fi} {(f+1) i \choose i}.
\ee
Recall now the well known statement $\sum_{i|n} \mu(i) = \left\{ \begin{smallmatrix}
1 & \ \textrm{for}\ n=1\\
0 & \ \textrm{for}\ n>1
\end{smallmatrix} \right.$,
which implies that for framing $f=0$ we get
\be
\Omega_n = -\frac{1}{n^2}  \sum_{i|n}  \mu(i) = \left\{ \begin{array}{cl}
-1 & \ \textrm{for}\ n=1\\
0 & \ \textrm{for}\ n>1
\end{array} \right.
\ee
This means that there is only one non-zero classical BPS state $\Omega_1=-1$, which is consistent with (\ref{Omega-numerical}) and having only one non-zero motivic Donaldson-Thomas invariant $\Omega_{1;0}=-1$, as mentioned below (\ref{P-C-1-x}).

Finally, the quantum curve is reduced to a differential equation upon the rescaling (\ref{rescale})
\be
( \partial_x - (-1)^f x) \widetilde{\psi}_f(x) = 0,
\ee
and for $f=0$ its solution $\widetilde{\psi}_{f=0}(x)$ is the simplest hypergeometric function (\ref{Frs})
\be
\widetilde{\psi}_{f=0}(x) = \, _r F_s[\, \cdot\, ;x] = e^x.   
\ee

%***************************************************************************************************

\subsection{Resolved conifold ($r=1, s=0$)}

\begin{figure}[h]
\begin{center}
\includegraphics[width=0.35\textwidth]{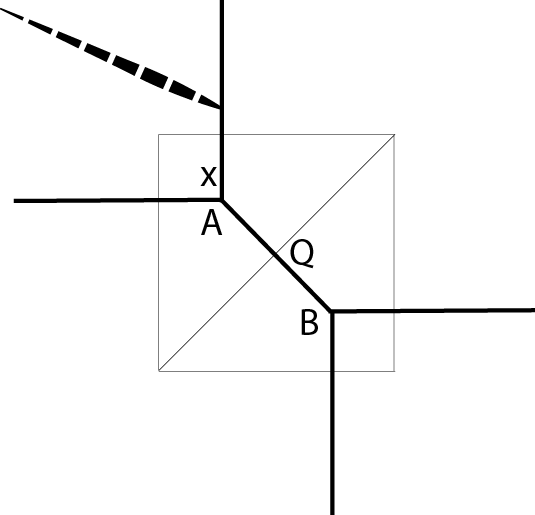} 
\caption{Conifold.}  \label{fig-strip-conifold}
\end{center}
\end{figure}

The second example we consider is the resolved conifold, whose toric diagram is shown in fig. \ref{fig-strip-conifold}. In this case the quiver matrix $C$ in (\ref{C-psi-f}), and the reduced matrix $C'$ introduced in (\ref{qHyper_quiver2}), take form
\be
  C_{r=1,s=0} = \begin{bmatrix}
    f+1 & 0 & 1 \\
    0 & 1 & 0 \\
    1 & 0 & 0
  \end{bmatrix} \qquad\quad
C'_{r=1,s=0} = \begin{bmatrix}
    f+1 &  1 \\
    1 & 0
  \end{bmatrix}   \label{C-Cprim-conifold}
\end{equation}
Note that $C$ is a different quiver (however it leads to the same generating function upon appropriate identification of parameters) than the one identified in \cite{Kucharski:2017ogk}, which had two vertices.

It is known that the brane amplitude in the resolved conifold geometry encodes colored HOMFLY-PT polynomials (without taking the extremal limit), and for special framing ($f=-1$ in our convention) the partition function can be resummed into a product of two quantum dilogarithms, which represent two BPS states. The brane partition function in this case depends just on two parameters, $x$ (brane modulus) and $\alpha=Q$ (conifold K{\"a}hler parameter). However, once considered as the quiver generating function, it arises from the identification of quiver generating parameters as in (\ref{psi-f-quiver}). In this case the quiver $C$ has three vertices, and its quiver generating function (\ref{P-C}) provides the refinement of the brane amplitude $\psi_f(x)$, and BPS invariants in particular. Indeed in framing $f=-1$, the general (without parameter identification) quiver generating series can be resummed to
\begin{equation}
  P_{C;f=-1}(x_1, x_2, x_3) = \frac{(q^{1/2 }x_2;q)_{\infty} (x_1 x_3;q)_{\infty}}{(x_1;q)_{\infty} (x_3;q)_{\infty}}.   \label{P-C-conifold-f-1}
\end{equation}
This means that there are four motivic Donaldson-Thomas invariants associated to this quiver 
\be
\Omega_{1,0,0;-1}=-1, \quad \Omega_{0,1,0;0}=-1,\quad  \Omega_{0,0,1;-1}=-1, \quad  \Omega_{1,0,1;-1}=1.   \label{BPS-conifold}
\ee 
As a consistency check, note that these invariants indeed satisfy the condition (\ref{Omega-constraint-3}).  On the other hand, the brane partition function $\psi_{f}(x)$ arises from the identification of parameters as in (\ref{psi-f-quiver}), and for $f=-1$ it reduces to the $q$-hypergeometric function (\ref{q-hyper}) with $\alpha=Q$
\begin{equation}
  \psi_{f=-1}(x) = P_C(x, q^{-1/2}\alpha, \alpha) =  \pphiq{1}{0}{\alpha}{\cdot}{q}{x} = \sum_{n=0}^{\infty} \frac{(\alpha;q)_{n}}{(q;q)_{n}} x^n = 
\frac{(\alpha x;q)_{\infty}}{ (x;q)_{\infty}} .   \label{psi-f-1-conifold}
\end{equation}
This is indeed well known product representation of the brane partition function in the conifold, which captures two BPS states that arise from the cancellation of the other two among those in (\ref{BPS-conifold}). The last equality in (\ref{psi-f-1-conifold}) is known as the $q$-binomial theorem.

In this example we can also identify difference operators (\ref{A-hat-PC}) that annihilate the quiver generating function (\ref{P-C-conifold-f-1}). For $f=-1$ we find that they take form
\begin{align}
  \hat{A}_1(\hat{x}_1, \hat{x}_2, \hat{x}_3, \hat{y}_1) &= (1-\hat{x}_1 \hat{x}_3) \hat{y}_1 -1 + \hat{x}_1,\\
  \hat{A}_2(\hat{x}_1, \hat{x}_2, \hat{x}_3, \hat{y}_2) &= (1- q^{1/2}\hat{x}_2)\hat{y}_2 - 1, \\
  \hat{A}_3(\hat{x}_1, \hat{x}_2, \hat{x}_3, \hat{y}_3) &= (1-\hat{x}_1 \hat{x}_3) \hat{y}_3 -1 + \hat{x}_1,
\end{align}
and in $q\to 1$ limit they reduce to classical partial $A$-polynomials
\begin{align}
  A_1(x_1, x_2, x_3, y_1) &= (1-x_1 x_3) y_1 -1 + x_1,\\
  A_2(x_1, x_2, x_3, y_2) &= (1-x_2)y_2 - 1, \\
  A_3(x_1, x_2, x_3, y_3) &= (1-x_1 x_3) y_3 -1 + x_1.
\end{align}
Because in the identification (\ref{psi-f-quiver}) it is just $x_1$ which is identified with the brane modulus $x$, it follows that an ordinary quantum A-polynomial that annihilates the unknot generating function (\ref{psi-f-1-conifold}) is identified simply with $\hat{A}_1$ and reads
\be
\widehat{A}(\hat x,\hat y) = \hat{A}_1(\hat{x}, q^{-1/2}\alpha, \alpha, \hat{y}) = (1-\alpha\hat{x}) \hat{y} -1 + \hat{x}.  \label{A-hat-conifold}
\ee
As a check, this indeed agrees with (\ref{A-A1-hats}), and in the classical limit this operator reduces to the well known conifold mirror curve
\be
A(x,y) = y -\alpha x y + x - 1 = 0.   \label{Axy-r1s0}
\ee
The coefficients of the series $y=\sum_{i,j} c_{i,j} x^i \alpha^j$ solving this equation follow from (\ref{y-c_i})
\be
c_{i,j} = \frac{(-1)^{i+j}}{1+j}\binom{j+1}{i}\binom{i}{j} = \left\{ \begin{array}{cl}
-1 & \ \textrm{for}\ j=i-1\\
1 & \ \textrm{for}\ j=i
\end{array} \right.,\qquad \textrm{for}\ i\geq 1,
\ee 
so that
\be
y=\sum_{i,j} c_{i,j} x^i \alpha^j = 1 - \sum_{i=1}^{\infty}  x^i (1 - \alpha)\alpha^{i-1} = \frac{1-x}{1-\alpha x},
\ee
which of course reproduces a direct solution of (\ref{Axy-r1s0}). One can also check that for $f=-1$ there are only two (associated to $C'$ in (\ref{C-Cprim-conifold})) non-zero classical BPS numbers (\ref{Omega-nlk-classical}), i.e. $\Omega_{1,0} = -1$ and $\Omega_{1,1} = 1$.

Finally consider the limit that turns (\ref{psi-f-1-conifold}) into an ordinary hypergeometric function. In the present example $1+s-r=0$, so the variable $x$ is not rescaled, and we identify $\alpha=q^a$ as in (\ref{rescale}). The quantum curve (\ref{A-hat-conifold}) reduces then to the hypergeometric equation (\ref{A-tilde-psi-f})
\be
(\partial_x-x\partial_x -a)\widetilde{\psi}_{f=-1}(x) = 0,
\ee
whose solution is the hypergeometric function $_1 F_0$, which indeed reproduces (\ref{PC-a1a2}) with $q^{a_1}\equiv q^{a-1/2}$ and $q^{a_2} \equiv q^a$
\be
\widetilde{\psi}_{f=-1}(x) = \pFq{1}{0}{a}{\cdot}{x} = (1-x)^{-a}.
\ee

%These equations are consistent with the classical $A$-polynomial
%\begin{equation}
% A(x_1, x_2, x_3, y) = (1-x_2)(1-x_1 x_3)^2 y - (1-x_1)(1-x_3),
%\end{equation}
%in the sense that if $y_1$, $y_2$ and $y_3$ solve $A_1$, $A_2$ and $A_3$ then $y = y_1 y_2 y_3$ solves $A$.  
%The quantum A-polynomial for the quiver $C_{r=1, s=0}$ can be also deduced from its classical limit. The result is
%\begin{equation}
%  \hat{A}(\hat{x}_1,\hat{x}_2,\hat{x}_3,\hat{y}) = (1 - q^{1/2} \hat{x}_2)(1 - (1+q) \hat{x}_1 \hat{x}_3 + q \hat{x}_1^2 \hat{x}_3^2)\hat{y} - (1-\hat{x}_1)(1-\hat{x}_3).
%\end{equation}

%***************************************************************************************************

\subsection{Resolution of $\mathbb{C}^3/\mathbb{Z}_2$ ($r=0, s=1$)}

\begin{figure}[h]
\begin{center}
\includegraphics[width=0.45\textwidth]{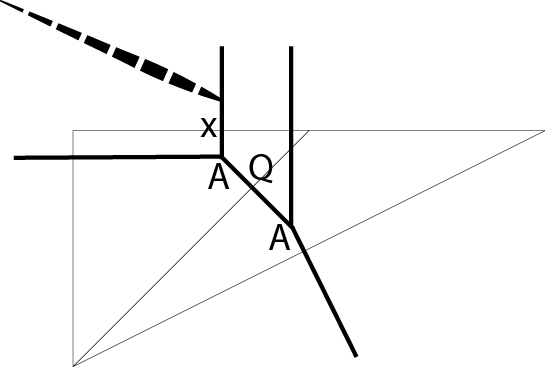} 
\caption{$\mathbb{C}^3/\mathbb{Z}^2$.}  \label{fig-strip-C3Z2}
\end{center}
\end{figure}

The next example we consider is the resolution of $\mathbb{C}^3/\mathbb{Z}_2$, see fig. \ref{fig-strip-C3Z2}, characterized by one K{\"a}hler parameter $\beta=Q$. In this case the corresponding quiver $C$ and its reduced counterpart $C'$ take form
\be
C_{r=0,s=1} = \begin{bmatrix}
    f+1 & 1 & 0 \\
    1 & 1 & 0 \\
    0 & 0 & 0
  \end{bmatrix}  \qquad\quad
C'_{r=0,s=1} = \begin{bmatrix}
    f+1 & 1 \\
    1 & 1 \\
  \end{bmatrix}   \label{C-Cprim-C3Z2}
\ee

In this case the framing that gives rise to the $q$-hypergeometric function is equal to $f=s-r=1$, and in this case the partition function cannot be represented as a product of a finite number of quantum dilogarithms. It only has a representation as an infinite product of quantum dilogarithms. The brane partition function in the representation (\ref{qHyper_quiver2}) takes form
\begin{equation}
\psi_{f=1}(x) =  \pphiq{0}{1}{\cdot}{\beta}{q}{x} = \frac{1}{(\beta;q)_{\infty}} P_{C'}(q^{-1}x, q^{-1/2}\beta).
\end{equation}
The product decomposition (\ref{PQx-Omega}) of the motivic generating function for the quiver $C'$ reads
\begin{align}
\begin{split}
  P_C'(x_1, x_2) &= \frac{(q^2 x_1^2;q) (q^{1/2}x_2;q) (q^{5/2} x_1^2 x_2;q) (q^{7/2} x_1^2 x_2;q) (q^3 x_1^2 x_2^2;q)^2 (q^4 x_1^2 x_2^2;q)  (q^5 x_1^2 x_2^2;q)}{(q x_1;q) (q^{3/2}x_1 x_2;q) (q^2 x_1 x_2^2;q) } + \\
  & \quad + \mathcal{O}(x_1^3,x_2^3) ,
\end{split}
\end{align}
which implies that several first motivic Donaldson-Thomas invariants associated to $C'$ are
\begin{align}
\begin{split}
  \Omega_{1,0;1} = -1, \quad \Omega_{0,1;0} = -1, \quad \Omega_{1,1;2} = 1, \quad \Omega_{2,0;3} = 1, \quad \Omega_{2,1;4} = -1,  \\
  \Omega_{2,1;6} = - 1, \quad \Omega_{1,2;3} = -1, \quad \Omega_{2,2;5} = 2, \quad \Omega_{2,2;7} = 1, \quad \Omega_{2,2;9} = 1.   \label{BPS-C3Z2}
\end{split}
\end{align}
After the identification of variables $x_1 = q^{-1} x$ and $x_2 = q^{-1/2} \beta$ the factor corresponding to the BPS number $\Omega_{0,1;0}$ cancel with the prefactor $(\beta;q)_{\infty}$.

In this example we can also identify partial quantum A-polynomials that annihilate $P_{C;f=1}(x_1,x_2,x_3)$; they take form
\begin{align}
  \hat{A}_1(\hat{x}_1, \hat{x}_2, \hat{x}_3, \hat{y}_1) &= (q\hat{x}_1 - q^{-1/2}\hat{x}_2) y_1^2 + (1+ q^{-1/2}\hat{x}_2) \hat{y}_1 - 1, \\
  \hat{A}_2(\hat{x}_1, \hat{x}_2, \hat{x}_3, \hat{y}_2) &= \hat{x}_1 \hat{y}_2^2 - (q^2 \hat{x}_2^2 - q^{3/2} \hat{x}_2 + q \hat{x}_1 + \hat{x}_1)\hat{y}_2 + (q \hat{x}_1 - q^{3/2} \hat{x}_2), \\
  \hat{A}_3(\hat{x}_1, \hat{x}_2, \hat{x}_3, \hat{y}_3) &= \hat{y}_3 + \hat{x}_3 - 1.
\end{align}
%The quantum $A$-polynomial is
%\begin{equation}
%  \hat{A}(\hat{x}_1, \hat{x}_2, \hat{x}_3, \hat{y}) = q \hat{x}_1 \hat{y}^2 + (1 - q^{1/2}\hat{x}_2 - q \hat{x}_3 + q^{3/2}\hat{x}_2 \hat{x}_3)\hat{y} - (1 - (1+q)\hat{x}_3 + q \hat{x}_3^2).
%\end{equation}
%The classical limit of which is
%\begin{equation}
%  A(x_1, x_2, x_3) = x_1 y^2 + (1 - x_2)(1-  x_3)y - (1 - x_3)^2.
%\end{equation}
The first of these operators, under the identification $\hat{x}_1 = q^{-1} \hat{x}$, $x_2 = q^{-1/2} \beta$, and $\hat{y}_1=\hat{y}$, reduces to the quantum A-polynomial that annihilates the brane partition function
\be
\hat{A}(\hat{x}, \hat{y}) = (\hat{x} - q^{-1}\beta) \hat{y}^2 + (1+ q^{-1}\beta) \hat{y} - 1,  \label{A-hat-C2Z2}
\ee
in agreement with (\ref{A-hat-psi-f}), and in $q\to 1$ limit we get the mirror curve
\be
A(x,y) = (x-\beta)y^2 + (1+\beta)y -1 = 0.
\ee
The solution of this equation for $y=y(x)$ follows from (\ref{y-c_i}) and of course it reproduces explicit solution of the quadratic equation
\be
y(x) = \sum_{i,j} \frac{(-1)^{3i+2j} i}{(i+j)(2i + j +1)} \binom{2i+j+1}{i}\binom{i+j}{j} x^i \beta^j = \frac{-1 - \beta + \sqrt{1 + 4x - 2\beta + \beta^2}}{2(x-\beta)}.
\ee
Furthermore, classical BPS numbers (\ref{Omega-nlk-classical}) (associated to $C'$ in (\ref{C-Cprim-C3Z2})) take form
\begin{align}
\begin{split}
  \Omega_{1, k} &= -1, -1, -1, -1, -1, \dots,\\
  \Omega_{2, k} &= 1, 2, 4, 6, 9, 12, \dots,\\
  \Omega_{3, k} &= -1, -5, -14, -31, -60, -105, \dots
\end{split}
\end{align}
etc., in agreement with (\ref{Omega-numerical}) and (\ref{BPS-C3Z2}).

On the other hand, in the limit that leads to a differential equation, in (\ref{rescale}) we need to rescale $x\to (q-1)^2 x$ and identify $\beta=q^b$. The quantum curve (\ref{A-hat-C2Z2}) reduces then to the hypergeometric equation (\ref{A-tilde-psi-f})
\be
(x \partial_x^2 + b\partial_x -1)\widetilde{\psi}_{f=1}(x) = 0,
\ee
whose solution is the hypergeometric function (\ref{Frs})
\be
\widetilde{\psi}_{f=1}(x) = \pFq{0}{1}{\cdot}{b}{x} = \sum_{n=0}^{\infty}  \frac{x^n}{n! (b)_n}.
\ee

%***************************************************************************************************

\subsection{Two K{\"a}hler parameters ($r=1, s=1$)}

\begin{figure}[h]
\begin{center}
\includegraphics[width=0.95\textwidth]{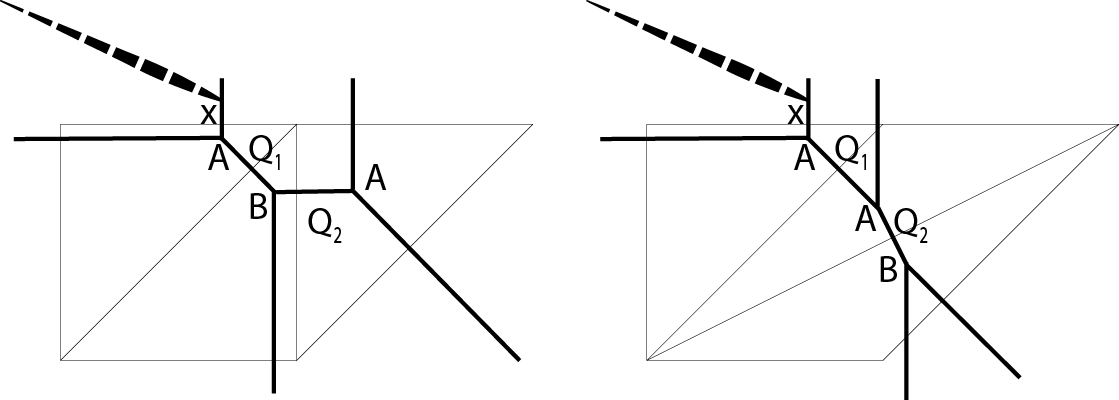} 
\caption{AAB, $_1\phi_1$.}  \label{fig-strip-1Phi1}
\end{center}
\end{figure}

As the next example we consider strip geometries with two K{\"a}hler parameters, for which a brane partition function is expressed in terms of the hypergeometric function $_1\phi_1$ with one argument $\alpha$ and one argument $\beta$. There are in fact two such manifolds, whose toric diagrams are shown in fig. \ref{fig-strip-1Phi1}. The first one includes two curves of type $(-1,-1)$, and was called a double-$\mathbb{P}^1$ in \cite{HSS}. The second one has one curve of type $(-2,0)$ and the other one of type $(-1,-1)$. These two geometries are related by the flop transition on $Q_2$. Even though brane partition functions for these two geometries are expressed in terms of the same function $_1\phi_1$, the identification of parameters is different in these two cases. Namely, in the former case, we set $\alpha=Q_1$ and $\beta=Q_1 Q_2$. In the latter case we set $\alpha=Q_1 Q_2$ and $\beta=Q_1$.

The quiver matrix (\ref{C-psi-f}) and its reduced form for these manifolds read
\be
  C_{r=1, s=1} = \begin{bmatrix}
    f+1 & 0 & 1 & 1 & 0 \\
    0 & 1 & 0 & 0 & 0 \\
    1 & 0 & 0 & 0 & 0 \\
    1 & 0 & 0 & 1 & 0 \\
    0 & 0 & 0 & 0 & 0
  \end{bmatrix}\qquad\quad
C'_{r=1, s=1} = \begin{bmatrix}
    f+1  & 1 & 1 \\
    1  & 0 & 0 \\
    1  & 0 & 1
  \end{bmatrix}   \label{quiver-r1s1}
\ee
For $f=0$ the motivic Donaldson-Thomas invariants associated to the quiver $C'$ take form
\begin{align}
\begin{split}
  \Omega_{1,0,0;0} &= -1, \quad \Omega_{0,1,0;-1} = -1, \quad \Omega_{1,0,1;1} = 1, \quad \Omega_{1,1,0;0} = 1, \\
  \Omega_{1,0,1;1} &= 1, \quad \Omega_{1,1,1;1} = -1, \quad \Omega_{1,0,2;2} = -1, \quad \Omega_{1,1,2;2} = -1, \\
  \Omega_{1,1,2;2} &= 2, \quad \Omega_{2,0,1;2} = -1, \quad \Omega_{2,0,2;3} = 1, \quad \Omega_{2,0,2;5} = 1,     \label{BPS-r1s1}
\end{split}
\end{align}
etc., and the brane partition function is expressed in terms of the $q$-hypergeometric function
\begin{equation}
\psi_{f=0}(x) =  \pphiq{1}{1}{\alpha}{\beta}{q}{x}.
\end{equation}

The partial quantum A-polynomial $\widehat{A}_1$ that annihilates the motivic generating function (\ref{P-C}) for the quiver (\ref{quiver-r1s1}) is
\begin{equation}
  \hat{A}_1(\hat{x}_1, \dots, \hat{x}_5, \hat{y}_1) = (-q^{1/2} \hat{x}_1 \hat{x}_3 + q^{-1/2}\hat{x}_4)\hat{y}_1^2 - ( 1 - q^{1/2} \hat{x}_1 + q^{-1/2}\hat{x}_4) \hat{y}_1 + 1.
\end{equation}
Changing parameters as in (\ref{qHyper_quiver}) this operator reduces to the quantum A-polynomial that annihilates $\psi_{f=0}(x)$
\be
\hat{A}(\hat{x},\hat{y}) = (-\alpha \hat{x} + q^{-1}\beta)\hat{y}^2 - ( 1 - \hat{x} + q^{-1}\beta) \hat{y} + 1,
\ee
in agreement with (\ref{A-hat-psi-f}) and (\ref{A-A1-hats}), and for $q\to 1$ it reduces to the mirror curve
\be
A(x,y) = (-\alpha x + \beta) y^2 - ( 1 - x + \beta) y + 1 = 0.
\ee
The solution of this equation for $y=y(x)$ again follows from (\ref{y-c_i})
\begin{align}
\begin{split}
y(x) &= \sum_{i,j,k} \frac{(-1)^j i}{(i+k)(i+j+k+1} \binom{i}{j}\binom{i+k}{k}\binom{i+j+k+1}{i}  x^i \alpha^j \beta^k =  \\
& = \frac{-1 + x - \beta + \sqrt{(-1+x-\beta)^2 + 4(\alpha x - \beta)}}{2(\alpha x - \beta)}.
\end{split}
\end{align}
Classical BPS invariants (\ref{Omega-nlk-classical}) in the case read
\begin{align}
\begin{split}
  \Omega_{1,0,k} &= 1, \quad \Omega_{1,1,k} = -1,  \quad \Omega_{1,j,k} = 0, \ \textrm{for}\  j \geq 2, \\
  \Omega_{2,0,k} &= 0, 1, 2, 4, 6, 9, \dots, \quad \Omega_{2,1,k} = -1, -3, -6, -10, -15, -21, \dots, \\
  \Omega_{2,2,k} &= 1, 2, 4, 6, 9, 12, \dots, \quad \Omega_{2,j,k} = 0, \ \textrm{for}  j \geq 3, \dots 
\end{split}
\end{align}
etc., in agreement with (\ref{Omega-numerical}) and (\ref{BPS-r1s1}).

Furthermore, rescaling $x\to (q-1)x$ and setting $\alpha=q^{a}$ and $\beta=q^b$ according to (\ref{rescale}), in the limit $q\to 1$, for $f=0$, the above quantum curve reduces to the hypergeometric equation
\begin{equation}
  (x \partial_x^2 + (b-x)\partial_x - a)\widetilde{\psi}_{f=0}(x),
\end{equation}
and the partition function reduces to the hypergeometric function
\be
\widetilde{\psi}_{f=0}(x) = \pFq{1}{1}{a}{b}{x} = \sum_{n=0}^{\infty}  \frac{(a)_n}{n! (b)_n} x^n.
\ee

%The full quantum $A$-polynomial
%\begin{align}
%  \hat{A}(\hat{x}_1, \dots, \hat{x}_5, \hat{y}) =\, &q^2 \hat{x}_1  \left(1  -q^{3/2} \hat{x}_2 - q^{1/2} \hat{x}_2+ q^2 \hat{x}_2^2\right) (\hat{x}_4-q \hat{x}_1 \hat{x}_3)\hat{y}^2\nonumber\\
%      &-\left(1-q^{1/2} \hat{x}_2\right) (1 - q \hat{x}_5) \left(1 + q^{3/2} \hat{x}_1 \hat{x}_3 - q^{1/2} (\hat{x}_4+\hat{x}_1 (1-\hat{x}_3))\right)\hat{y} \nonumber \\
%  &-q \hat{x}_3 \hat{x}_5^2+(q+1) \hat{x}_3 \hat{x}_5+q \hat{x}_5^2-(q+1) \hat{x}_5-\hat{x}_3+1.
%\end{align}
%In the classical limit
%\begin{align}
%  A(x_1, \dots, x_5, y) =\, & x_1  \left( 1- x_2\right)^2 (x_4- x_1 x_3)y^2\nonumber\\
%      &-\left(1- x_2\right) ( 1-x_5) \left(1 - x_4 - x_1 +  2x_1 x_3\right)y \nonumber \\
%  &- x_3 x_5^2+2 x_3 x_5+ x_5^2-2 x_5-x_3+1.
%\end{align}

%***************************************************************************************************

\subsection{Two other K{\"a}hler parameters ($r=2$, $s=0$)}

\begin{figure}[h]
\begin{center}
\includegraphics[width=0.45\textwidth]{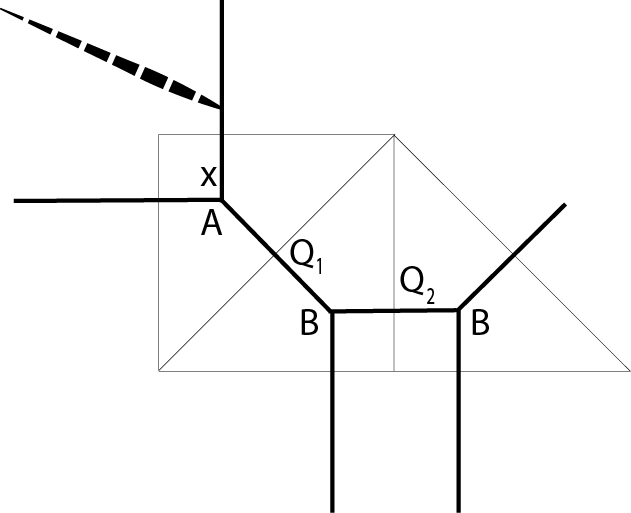} 
\caption{ABB, $_2\phi_0$.}  \label{fig-strip-2Phi0}
\end{center}
\end{figure}

Now we consider another strip geometry with two K{\"a}hler parameters, shown in fig. \ref{fig-strip-2Phi0}. In this case we identify parameters as $\alpha_1=Q_1$ and $\alpha_2 = Q_1 Q_2$. We find that the full and reduced quiver matrices take form
\begin{equation}
  C = \begin{bmatrix}
    f+1  & 0 & 1  & 0 & 1 \\
    0 & 1 & 0 & 0 & 0 \\
    1  & 0 & 0  & 0 & 0 \\
    0 & 0 & 0 & 1 & 0 \\
    1  & 0  & 0 & 0 & 0
  \end{bmatrix}\qquad \quad
  C' = \begin{bmatrix}
    f+1  & 1  & 1 \\
    1  & 0  & 0 \\
    1  & 0  & 0
  \end{bmatrix}
\end{equation}
For framing $f=s-r=-2$ the motivic Donaldson-Thomas invariants for the series $P_{C'}(x_1, x_2, x_3)$ associated to the quiver $C'$ take form
\begin{align}
\begin{split}
  \Omega_{1,0,0;-2} &= -1, \quad \Omega_{0,1,0;-1} = -1, \quad \Omega_{0,0,1;-1} = -1, \quad \Omega_{1,1,0;-2} = 1, \quad \Omega_{1,0,1,-2} = 1, \\
 \Omega_{1,1,1;-2} &= -1, \quad \Omega_{2,0,0;-5} = -1, \quad \Omega_{2,1,0;-5} = 1, \quad
  \Omega_{2,0,1;-5} = 1, \quad \Omega_{2,1,1;-5} = -1, 
\end{split}
\end{align}
%After specialization of the variables, factors with DT numbers $\omega_{0,1,0:-1}$ and $\omega_{0,1:-1}$ cancel pairwise with factors $(\alpha_1; q)_{\infty}$ $(\alpha_2; q)_{\infty}$.
etc., which are identified with BPS number upon the identification of parameters (\ref{qHyper_quiver}). For $f=-2$ the brane partition function takes form of the $q$-hypergeometric function
\begin{equation}
 \psi_{f=-2}(x)  = (\alpha_1; q)_{\infty} (\alpha_2; q)_{\infty} \,P_{C'}(q^{1/2}x, \alpha_1, \alpha_2) = \pphiq{2}{0}{\alpha_1, \alpha_2}{\cdot}{q}{x}.
\end{equation}

The quantum curve can be easily derived from the general expression (\ref{A-hat-psi-f}), as in earlier examples. In the classical limit it reduces to the mirror curve
\be
A(x,y) = (1 - y)y + x (1 - \alpha_1) (1-\alpha_2 y) = 0,
\ee
and its solution for $y=y(x)$ again follows form (\ref{y-c_i})
\begin{align}
\begin{split}
y(x)&=\sum_{i,j,k} \frac{(-1)^{j+k} (j+k - 2i +2)_{i-1}}{i!} \binom{i}{j}\binom{i}{k} x^i \alpha_1^j\alpha_2^k = \\
& = \frac{-1 + \alpha_1 x + \alpha_2 x - \sqrt{(1- \alpha_1 x - \alpha_2 x)^2 - 4 x(\alpha_1 \alpha_2 x - 1)}}{2(\alpha_1 \alpha_2 x - 1)}.
\end{split}
\end{align}

%***************************************************************************************************

\subsection{Resolution of $\mathbb{C}^3/\mathbb{Z}_3$ ($r=0, s=2$)}

\begin{figure}[h]
\begin{center}
\includegraphics[width=0.55\textwidth]{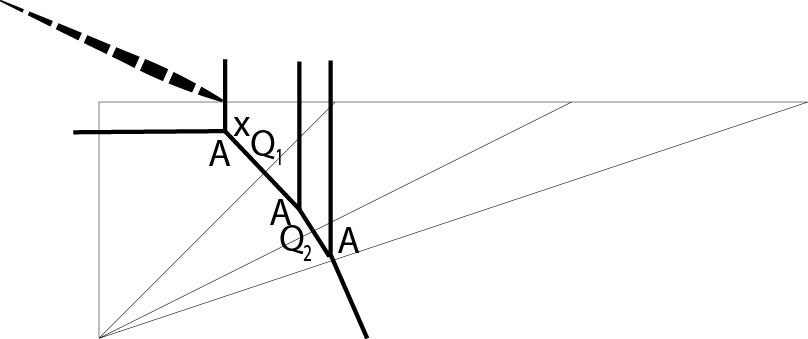} 
\caption{The resolution of $\mathbb{C}^3/\mathbb{Z}_3$.}  \label{fig-strip-C3Z3}
\end{center}
\end{figure}

Another example is the resolution of $\mathbb{C}^3/\mathbb{Z}_3$, shown in fig. \ref{fig-strip-C3Z3}. In this case parameters are identified as $\beta_1=Q_1$ and $\beta_2=Q_1 Q_2$. The reduced quiver matrix takes form
\begin{equation}
  C_{r=0,s=2} = \left[\begin{matrix}
    f+1 & 1 & 0& 1 &0\\
    1 & 1 & 0& 0 & 0\\
    0 & 0 & 0 & 0 & 0\\
    1 & 0 & 0& 1 &0 \\
    0 & 0 & 0 & 0 & 0
    \end{matrix}\right]\qquad\qquad
  C'_{r=0,s=2} = \left[\begin{matrix}
    f+1 & 1 & 1 \\
    1 & 1 & 0 \\
    1 & 0 & 1
    \end{matrix}\right]
\end{equation}
For framing $f=s-r=2$ the brane partition function takes form of the $q$-hypergeometric function
\begin{equation}
\psi_{f=2} = \frac{1}{(\beta_1; q)_{\infty} (\beta_2; q)_{\infty}} \,P_{C'}(q^{-3/2}x, q^{-1/2}\beta_1, q^{-1/2}\beta_2) =  \pphiq{0}{2}{\cdot}{\beta_1, \beta_2}{q}{x}.
\end{equation}
The quantum curve is again derived from the general expression (\ref{A-hat-psi-f}). In the classical limit for $f=2$ it reduces to the mirror curve
\be
A(x,y) = (1-y)(1 - \beta_1 y)(1-\beta_2 y) + x y^3 = 0,
\ee
and the solution of this cubic equation for $y=y(x)$ also follows form (\ref{y-c_i})
\begin{align}
\begin{split}
y(x) &= \sum_{i,j,k} \frac{i^2}{(1 + 3i + j + k)(i+j)(i+k)}\binom{1 + 3i + j +k}{i}\binom{i+j}{j}\binom{i+k}{k} x^i \beta_1^j \beta_2^k = \\
& = 1 + \frac{1}{(\beta_1 - 1)(\beta_2 - 1)}x + \frac{3 -2 (\beta_1+\beta_2) + \beta_1\beta_2}{(\beta_1 - 1)^3(\beta_2 - 1)^3} x^2 + \ldots
\end{split}
\end{align}

%***************************************************************************************************

\subsection{Three K{\"a}hler parameters and $q$-hypergeometric function ($r=2$, $s=1$)}

\begin{figure}[h]
\begin{center}
\includegraphics[width=0.45\textwidth]{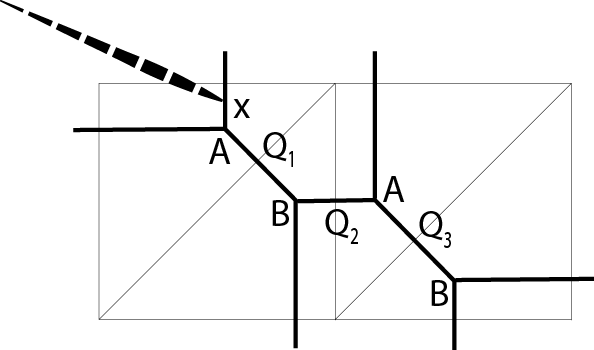} 
\caption{Triple-$\mathbb{P}^1$ geometry.}  \label{fig-triple}
\end{center}
\end{figure}

Finally we consider the geometry with three K{\"a}hler parameters, such that -- apart from the first vertex of type A -- another $r=2$ vertices are of type B and $s=1$ vertex is of type A. There are three manifolds of this type, with vertices distributed in the order ABAB, AABB, or ABBA. In all these cases the brane partition function can be written in the form (\ref{psi-f-quiver}), with the corresponding quiver matrix and its reduced form given by
\begin{equation}
  C_{r=2,s=1} = \begin{bmatrix}
    f+1 & 0 & 1 & 0 & 1 & 1 & 0 \\
    0 & 1 & 0 & 0 & 0 & 0 & 0 \\
    1 & 0 & 0 & 0 & 0 & 0 & 0 \\
    0 & 0 & 0 & 1 & 0 & 0 & 0 \\
    1 & 0 & 0 & 0 & 0 & 0 & 0 \\
    1 & 0 & 0 & 0 & 0 & 1 & 0 \\
    0 & 0 & 0 & 0 & 0 & 0 & 0 \\
  \end{bmatrix}\qquad\quad
  C'_{r=2,s=1} = \begin{bmatrix}
    f+1 & 1 & 1 & 1 \\
    1 & 0 & 0 & 0 \\
    1 & 0 & 0 & 0 \\
    1 & 0 & 0 & 1 \\
  \end{bmatrix}
\end{equation}
These three cases differ by the assignment of K{\"a}hler parameters, which respectively take the following form:
\begin{align}
\begin{split}
\textrm{ABAB}:&\qquad \alpha_1 = Q_1,\ \alpha_2 = Q_1Q_2Q_3,\ \beta_1 = Q_1Q_2, \\
\textrm{AABB}:&\qquad \alpha_1 = Q_1Q_2,\ \alpha_2 = Q_1Q_2Q_3,\ \beta_1 = Q_1,\\
\textrm{ABBA}:&\qquad \alpha_1 = Q_1,\ \alpha_2 = Q_1Q_2,\ \beta_1 = Q_1Q_2Q_3.
\end{split}
\end{align}
As one example, the geometry with vertices ABAB is shown in fig. \ref{fig-triple}.

In all these cases, in the framing $f=s-r=-1$, the partition function (\ref{q-hyper}) reduces to the (proper, not ``generalized'') $q$-hypergeometric function, which can also be expressed in terms of the motivic generating function for the reduced quiver
\begin{equation}
\psi_{s-r}(x)  = \frac{(\alpha_1; q)_{\infty} (\alpha_2; q)_{\infty}}{(\beta_1;q)} \,P_{C'_{r=2,s=1}}(x, \alpha_1, \alpha_2, q^{-1/2}\beta_1) =  \pphiq{2}{1}{\alpha_1, \alpha_2}{\beta_1}{q}{x}.
\end{equation}
The non-zero motivic Donaldson-Thomas invariants associated to the latter generating series $P_{C'_{r=2,s=1}}(x_1, x_2, x_3, x_4)$, for $f=-1$, take form
\begin{align}
\begin{split}
  \Omega_{0,0,0,1;0} &= -1, \quad \Omega_{0,0,1,0;-1} = -1, \quad \Omega_{0,1,0,0;-1} = -1, \quad \Omega_{1,0,0,0;-1} = -1, \\
  \Omega_{1,0,0,1;0} &= 1, \quad \Omega_{1,0,1,0;-1} = 1, \quad \Omega_{1,1,0,0;-1} = 1, \quad \Omega_{1,1,0,1;0} = -1, \\
  \Omega_{1,0,1,1;0} &= 1, \quad \Omega_{1,1,1,0;-1} = -1, \quad \Omega_{1,1,1,1,;0} = 1, 
  \end{split}
\end{align}
etc. After rescaling (\ref{rescale}) and taking the limit $q\to 1$, for $f=-1$ the brane partition function reduces to the ordinary hypergeometric function  (\ref{Frs})
\be
\widetilde{\psi}_{f=-1}(x) = \pFq{2}{1}{a_1, a_2}{b_1}{x} = \sum_{n=0}^{\infty}  \frac{(a_1)_n (a_2)_n}{n! (b_1)_n} x^n.
\ee

%***************************************************************************************************
%***************************************************************************************************
%***************************************************************************************************
%***************************************************************************************************

\appendix

\section{Quiver generating functions and motivic Donaldson-Thomas invariants}   \label{sec-app}

In this appendix we compare our notation to that of Efimov in \cite{efimov2012}, who associates to a symmetric quiver, determined by a symmetric matrix $M$ with non-negative entries $M_{ij}\geq 0$, the motivic generating series of the form
\begin{equation}
  P_M^{\rm E}(x_1, \dots, x_m;q) = \sum_{d_1, \dots, d_m \geq 0} \frac{\left(-q^{1/2}\right)^{\sum_{i} d_i^2 - \sum_{i,j} M_{i,j} d_i d_j}}{(q;q)_{d_1} \cdots (q;q)_{d_m}} x_1^{d_1} \cdots x_m^{d_m}.  \label{P-C-Efimov}
\end{equation}
It is proved in \cite{efimov2012} that the above series encodes non-negative motivic Donaldson-Thomas invariants $c_{d_1, \dots, d_m; k}\geq 0$, which are determined by the factorization 
\begin{equation}
  P_M^{\rm E}(x_1, \dots, x_m;q) = \prod_{(d_1, \dots, d_m)> 0} \prod_{k \in \mathbb{Z}}\left(q^{k/2} x_1^{d_1}\cdots x_m^{d_m};q\right)_{\infty}^{(-1)^{k-1}c_{d_1, \dots, d_m; k}}.     \label{P-C-Efimov-prod}
\end{equation}
%For $m_{i,j}$ negative it seems that $c_{d_1, \dots, d_m; k}$ are still integers, however, might be now also negative.

Let us compare these definitions to our conventions (\ref{P-C}) and (\ref{PQx-Omega}). At first, one might wish to identify our matrix with entries $C_{i,j}$ with $\delta_{i,j} - M_{i,j}$ in (\ref{P-C-Efimov}). However, as our $C_{i,j}$ are positive (with positive $C_{1,1}$ at least for appropriately chosen framing $f$), this would mean that $M_{i,j}$ are not all positive, and in this case the proof in \cite{efimov2012} would not work (if some $M_{i,j}$ are negative, then exponents arising in the factorization (\ref{P-C-Efimov-prod}) are still integer, however not necessarily non-negative).

Nonetheless, we can relate to each other the generating functions (\ref{P-C}) and (\ref{P-C-Efimov}), and corresponding integer invariants, by inverting $q$. Indeed, denoting $|d|=d_1+\ldots +d_m$, we get
\begin{align}
\begin{split}
& P_C(x_1,\ldots,x_m;q^{-1})=\sum_{d_1,\ldots,d_m} \frac{(-q^{1/2})^{\sum_{i,j=1}^m (\delta_{i,j}-C_{i,j})d_id_j}}{(q;q)_{d_1}\cdots(q;q)_{d_m}} (q^{1/2}x_1)^{d_1}\cdots (q^{1/2}x_m)^{d_m} = \\
& \quad = P^{\rm E}_{\rm C}(q^{1/2}x_1, \dots, q^{1/2}x_m;q)   = \prod_{(d_1, \dots, d_m)> 0} \prod_{k \in \mathbb{Z}}\left(q^{(k+|d|)/2} x_1^{d_1}\cdots x_m^{d_m};q\right)_{\infty}^{(-1)^{k-1}c_{d_1, \dots, d_m; k}},       \label{P-C-compare}
\end{split}
\end{align}
now with non-negative integers $c_{d_1, \dots, d_m; k}$. Let us now compare these $c_{d_1, \dots, d_m; k}$ to our $\Omega_{d_1,\ldots,d_m;j}$, by relating the product expansion in (\ref{P-C-compare}) to that in (\ref{PQx-Omega})
\begin{align}
\begin{split}
P_C(x_1,\ldots,x_m;q) & = 
\prod_{(d_1, \dots, d_m)> 0} \prod_{k \in \mathbb{Z}}\left(q^{-(k+|d|)/2} x_1^{d_1}\cdots x_m^{d_m};q^{-1}\right)_{\infty}^{(-1)^{k-1}c_{d_1, \dots, d_m; k}} = \\
& = \prod_{(d_1,\ldots,d_m)\neq 0} \prod_{j\in\mathbb{Z}}  \Big( q^{(j+1)/2}  x_1^{d_1}\cdots x_m^{d_m} ;q  \Big)_{\infty}^{(-1)^{j+1}\Omega_{d_1,\ldots,d_m;j}}.
\end{split}
\end{align}
The relation between $c_{d_1, \dots, d_m; k}$ and $\Omega_{d_1,\ldots,d_m;j}$ can be found by matching powers of~$x_i$'s. Assume that we have matched the integers up to a certain power, and we wish now to match the next coefficient at ${\bf x}^{\bf d}\equiv x_1^{d_1}\cdots x_m^{d_m}$, where $d_j$ is already increased by 1. We can expand the corresponding quantum dilogarithms, and in the leading order we find
\begin{align}
\begin{split}
  (q^{-k/2 - |d|/2} {\bf x}^{\bf d}; q^{-1})_{\infty}^{(-1)^{k-1} c_{d_1, \dots, d_m; k}} &= 1 + (-1)^{k-1}c_{d_1, \dots, d_m; k} \frac{q^{-k/2 - |d/2| + 1}}{1 - q} {\bf x}^{\bf d} + \dots,\\
  (q^{(j+1)/2} {\bf x}^{\bf d}; q)_{\infty}^{(-1)^{j+1}\Omega_{d_1, \dots, d_m; j}} &= 1 - (-1)^{j+1}\Omega_{d_1, \dots, d_m; j} \frac{q^{(j+1)/2}}{1 - q} {\bf x}^{\bf d} + \dots.
\end{split}
\end{align}
As we assumed that all lower orders are already matched, these two terms must be matched on their own
\begin{equation}
  (-1)^{k-1}c_{d_1, \dots, d_m; k} q^{-(k + |d|)/2 + 1} = -(-1)^{j+1}\Omega_{d_1, \dots, d_m; j}q^{(j+1)/2}.
\end{equation}
Fixing $j$ so that the powers of $q$ are equal we find
\begin{equation}
  j = - k  - |d| + 1,
\end{equation}
and in consequence
\begin{equation}
 c_{d_1, \dots, d_m; k} =  (-1)^{|d|} \Omega_{d_1, \dots, d_m; -k -|d|+1} .  \label{c-Omega}
\end{equation}
%Or equivalently
%\begin{equation}
%  (-1)^{|d|+1}\Omega_{d_1, \dots, d_m; j} =  c_{d_1, \dots, d_m; -j +1 - |d|}. \label{relation_DT2}
%\end{equation}
Note that in all examples considered in the main text, for which all entries of the matrix $C$ or $C'$ are non-negative, multiplying the $\Omega_{d_1,\ldots,d_m;j}$ by $(-1)^{|d|}$  indeed produces non-negative integers.  Moreover, from (\ref{c-Omega}) we deduce that
\be
\Omega_{d_1, \dots, d_m} = \sum_j (-1)^j \Omega_{d_1, \dots, d_m;j} = -\sum_k (-1)^k c_{d_1, \dots, d_m;k} \equiv -c_{d_1,\ldots,d_m}.
\ee
Therefore numerical Donaldson-Thomas invariants defined either as in (\ref{Omega-numerical}) in terms of $\Omega_{d_1, \dots, d_m;j}$, or analogously in terms of non-negative $c_{d_1, \dots, d_m;k}$ introduced in (\ref{P-C-compare}), differ only by the overall sign.

%The classical limit can be written now in two ways
%\begin{align}
%  y(x_1, \dots, x_m) = \lim_{q\rightarrow 1} \frac{P_{\bar{C}}(q x_1, \dots, q x_m)}{P_{\bar{C}}(x_1, \dots, x_m)} &= \prod_{(d_1, \dots, d_m)> 0} \left(1 - x_1^{d_1} \cdots x_m^{d_m} \right)^{|d| \sum_k (-1)^{k-1} c_{d_1, \dots, d_m; k}} \nonumber \\
%  &= \prod_{(d_1, \dots, d_m)> 0} \left(1 - x_1^{d_1} \cdots x_m^{d_m} \right)^{- |d| \sum_j (-1)^{j+1} \Omega_{d_1, \dots, d_m;j}}
%\end{align}

%************************************************************************************
%************************************************************************************
%************************************************************************************

\acknowledgments{We thank Tobias Ekholm, Sergei Gukov, Piotr Kucharski, H{\'e}lder Larragu{\'i}vel, Chiu-Chu Melissa Liu, Pietro Longhi, Marko Sto$\check{\text{s}}$i$\acute{\text{c}}$, Cumrun Vafa, Johannes Walcher, and Don Zagier for inspiring discussions. P.S. thanks Aspen Center for Physics, Simons Center for Geometry and Physics, International Centre for Theoretical Sciences in Bangalore, Banff International Research Station, and Kavli Institute for Theoretical Physics at the University of California Santa Barbara, where parts of this work were done, for hospitality. This work is supported by the ERC Starting Grant no. 335739 \emph{``Quantum fields and knot homologies''} funded by the European Research Council under the European Union's Seventh Framework Programme, and the Foundation for Polish Science. M.P. acknowledges the support from the National Science Centre through the FUGA grant 2015/16/S/ST2/00448}.

%************************************************************************************
%************************************************************************************
%************************************************************************************

\newpage

\bibliographystyle{JHEP}
\bibliography{abmodel}

\end{document}